\documentclass[aip, amsmath,amssymb,reprint]{revtex4-1}

\usepackage{graphicx}
\usepackage{dcolumn}
\usepackage{bm}
\usepackage[mathlines]{lineno}

\usepackage[utf8]{inputenc}
\usepackage[T1]{fontenc}
\usepackage{mathptmx}
\usepackage{etoolbox}

\usepackage{xcolor}
\usepackage{booktabs}
\usepackage{multirow}

\makeatletter
\def\@email#1#2{%
 \endgroup
 \patchcmd{\titleblock@produce}
  {\frontmatter@RRAPformat}
  {\frontmatter@RRAPformat{\produce@RRAP{*#1\href{mailto:#2}{#2}}}\frontmatter@RRAPformat}
  {}{}
}%
\makeatother
\begin{document}


\title[Resonant frequency control for kinetic inductance detectors]{In-situ control of the resonant frequency of kinetic inductance detectors with multiplexed readout}
\author{M. Rouble}
\affiliation{Department of Physics and Trottier Space Institute, McGill University, Montreal, QC, H3A 2T8, Canada}
\email{maclean.rouble@mail.mcgill.ca}
\author{M. Adami\v{c}}
\affiliation{Department of Physics and Trottier Space Institute, McGill University, Montreal, QC, H3A 2T8, Canada}

\author{P. S. Barry}
\affiliation{School of Physics and Astronomy, Cardiff University, Cardiff, CF24 3AA, UK}

\author{K. R. Dibert}
\affiliation{Department of Physics, University of Chicago, Chicago, IL, USA}
\affiliation{Kavli Institute for Cosmological Physics, University of Chicago, Chicago, IL, USA}

\author{M. Dobbs}
\affiliation{Department of Physics and Trottier Space Institute, McGill University, Montreal, QC, H3A 2T8, Canada}

\author{K. Fichman}
\affiliation{Department of Physics, University of Chicago, Chicago, IL, USA}
\affiliation{Kavli Institute for Cosmological Physics, University of Chicago, Chicago, IL, USA}

\author{J. Montgomery}
\affiliation{Department of Physics and Trottier Space Institute, McGill University, Montreal, QC, H3A 2T8, Canada}

\author{G. Smecher}%
\affiliation{t0.technology inc, Montreal, QC, Canada}

\date{\today}

\begin{abstract}

Large multiplexing factors are a primary advantage of kinetic inductance detectors (KIDs), but the implementation of high density arrays still presents significant challenges.
Deviations between designed and achieved resonant frequencies are common, and differential loading and responsivity variation across an array may lead to dynamic inter-resonator interactions. It is therefore valuable to be able to both set and maintain the resonant frequency of a KID in situ, using the readout system. 
We show that it is possible to alter the resonant frequency of the devices by multiple linewidths through the application of readout current, and establish a new stable operational bias point at the driven frequency by making use of the hysteretic bistability commonly seen as bifurcation in frequency-domain measurements. 
We examine this interaction using a readout tone at fixed frequency positioned near or within the unbiased resonant bandwidth.
Development of a control methodology based on this principle remains in an early stage, but a foundational step is understanding the interaction of the readout current with the resonator, in particular its influence on the resonant frequency.
In this work, we study conventional KIDs with no physical isolation from the substrate, so we posit that the readout current primarily interacts with the resonator via non-thermal mechanisms, resulting in a predominantly reactive response. 
This behaviour is reproduced by a simple lumped-element circuit model of the resonance and readout system, providing a straightforward framework for analysis and interpretation.
This demonstration is an important early step in the development of techniques which seek to dynamically alter the resonant frequencies of conventional KID arrays, and sets the stage for fast active resonant frequency control under operational conditions.

\end{abstract}

\maketitle

\section{\label{sec:level1}Introduction}

Modern telescopes demand increasingly large and dense detector arrays, and are exploring novel experiment architectures. Kinetic inductance detectors\cite{mazin2002,day2003} (KIDs) offer high multiplexing densities and, compared with the current state-of-the-art transition edge sensor arrays, simplified fabrication and greatly reduced cryogenic hardware complexity.
These attributes make them an attractive choice for these applications, with their potential extending beyond imaging arrays to more complex on‑chip devices, such as filterbank spectrometers.\cite{endo2019,karkare2021}

The development of these large‑scale KID arrays is not without challenges. The frequency placement of resonators along a feedline is highly sensitive to fabrication systematics and variations in material properties. This leads to scatter in the achieved versus designed resonant frequencies, and consequently, resonances that overlap with one another. Recent advances in post‑fabrication tuning, such as by physically trimming the resonator capacitors, have shown that this scatter can be corrected at scale,\cite{vissers2024} but this process adds complexity and lengthens the amount of time required to go from design to operational array.

Further challenges arise during operation due to variations in responsivity and loading across resonators sharing a feedline: certain resonators will shift in frequency more than others in response to a signal, potentially colliding with their neighbours. 
This may be especially problematic for a spectrometer, where adjacent detectors are more likely to experience radically different loading levels due to narrowband line emission. 
When collided, the two resonances and their corresponding readout tones will interfere. If one resonance moves past another, it may not be possible to reestablish which is which. Avoiding this issue typically means either deliberately lowering responsivity in the design of the sensors or reducing the resonator density in frequency space (the multiplexing factor).

The ability to adjust resonant frequencies in situ would provide a solution to these issues, with the potential to improve array yield and increase multiplexing density. Although primarily intended to sense changes in optical or thermal loading, KIDs also demonstrate sensitivity to the low-energy photons of the readout current. This readout sensitivity is most commonly noted in the apparent distortion incurred in the resonance shape when sweeping a tone of sufficient amplitude across the resonant bandwidth. As the readout tone is swept across the resonance, the impedance it probes changes with frequency, shunting varying amounts of current through the resonance, and the resonance moves in response. Because a conventional sequential frequency sweep measurement records data at only one frequency at a time, the simultaneous motion of the resonance and readout tone complicates the view of the underlying resonator transfer function.

The strength of this sensitivity to the readout current suggests an avenue for resonant frequency control. To date, most if not all explorations of resonator current response have used either direct current\cite{vissers2015} or a readout tone which is swept in frequency across the resonator bandwidth.\cite{deVisser2010,Swenson2013} The former typically requires a device which is purpose‑built, in order to channel the current into the resonant circuit. The latter method, although broadly applicable, can obscure the effects of the readout current due to the dynamic nature of the measurement.

Recent work has demonstrated that this readout current sensitivity may be harnessed to both perturb and actively maintain the resonant frequency of a non‑thermally‑isolated KID during changes in loading.\cite{rouble2024} A system capable of stabilizing the resonators in place during operation would eliminate concerns about dynamic interactions between resonators. It also suggests the possibility of increasing the base array yield by separating resonators which are collided in their relaxed states, reducing the need to trim. However, these techniques are still in the very early stages of development, requiring further study of the readout‑resonator interaction.

To this end, we make use of a static‑frequency readout tone to inject current into the resonance and evaluate the resulting change in its resonant frequency.
We build a lumped‑element circuit model of the resonator, which we use to predict and analyze the impact of the readout current on the resonator. We show that we can manipulate the resonant frequency of the device over a wide bandwidth and establish a new stable bias point more than a linewidth away from the relaxed resonant frequency, and that this behaviour is readily reproduced using the circuit model.

\section{Current-dependent lumped-element circuit model}
\subsection{Transfer function}

\begin{figure}[h]
    \centering
    \includegraphics[width=\linewidth]{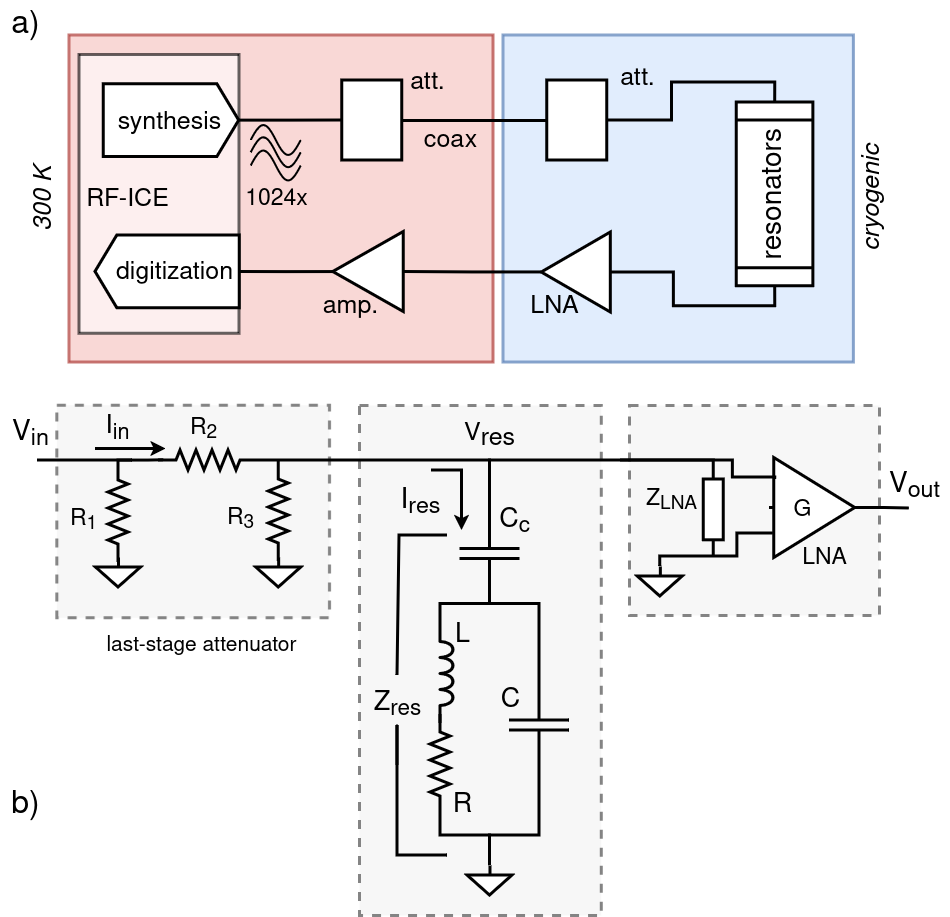}
    \caption{(a) Diagram of measurement setup. The RF-ICE\cite{rouble2022} readout platform provides up to 1024 sinusoidal readout tones, whose amplitudes and frequencies can be independently and dynamically updated. These are transported by coaxial cables through several attenuation stages to the resonator array, located within a BlueFors LD250S dilution refrigerator. The amplitude of the returning signals is conditioned by a cryogenic low-noise amplifier (LNA) and one or more stages of warm amplification, before being digitized by the RF-ICE hardware. (b) Circuit model of a single lumped element KID and surrounding readout electronics. Modeling the resonator in this simple circuit framework facilitates the study of the impact of readout current on its parameters, including predicting the current-dependent resonant frequency.}
    \label{fig:circuit_model}
\end{figure}

The amount of readout current entering the resonator is determined by its impedance and the impedances of the rest of the circuit as depicted in Fig. \ref{fig:circuit_model}b, evaluated at the frequency of the readout tone, $f$.
Following a standard lumped-element circuit analysis, we model the resonator as an inductance, $L = L_k + L_g$ (with $L_k$ and $L_g$ the kinetic and geometric components, respectively), and a small series real impedance $R$ arising from the real component of the complex conductance, in parallel with a capacitor, $C$:

\begin{equation}
    Z_{RLC} = \left[ (j2\pi f L + R)^{-1} + \left(\frac{1}{j2\pi f C}\right)^{-1}\right]^{-1}
\end{equation}

\noindent all of which is in series with a coupling capacitance, $C_c$, such that the total resonant branch has impedance

\begin{equation}\label{eq:Zres}
    Z_{res} = \frac{1}{j 2\pi f C_c} + Z_{RLC} \quad { .}
\end{equation}

\noindent The resonator is in parallel with the input impedance of the cryogenic low-noise amplifier, $Z_{LNA}$, and the resistance, $R_3$, of the last-stage attenuator. This parallel network forms the load on the last-stage attenuator. Approximating that the voltage at the point $V_{in}$ is fixed, the transfer function for this system from this point to the output of the LNA is:

\begin{equation}\label{eq:transfer_function}
    V_{out} = V_{res} G_{LNA} = V_{in} \frac{Z_{||}}{Z_{||} + R_2} G_{LNA} \quad { ,}
\end{equation}

\noindent with $Z_{||}$ the total impedance of the parallel network including the resonator ($Z_{res}$, $Z_{LNA}$, $R_3$, and any other resonators with resonant frequencies near to that of the device of interest, not included in the system diagram in Fig. \ref{fig:circuit_model}).

The readout current flowing through the resonator may be calculated using a current divider:

\begin{equation}\label{eq:Ires}
    I_{res} = I_{in} \frac{Z_{||}}{Z_{res}}
\end{equation}

\noindent with the input current to the network $I_{in} = V_{in} / (Z_{||} + R_2)$.

From this, we obtain the current flowing through the inductor, $I_L$, as:

\begin{equation}\label{eq:I_L}
	I_L = I_{res} \frac{Z_{RLC}}{j2\pi f L + R}  \quad { .}
\end{equation}

\paragraph{Zero-current impedance.}

When no current flows through the resonator, the zero-current kinetic inductance, $L_k (0)$, is derived from the geometry and material properties of the superconductor, via the dirty limit surface impedance\cite{henkels1977,deVisser2014}:

\begin{equation}
    Z_s = \sqrt{\frac{j 2 \pi f \mu_0}{\sigma}} \left(\mathrm{tanh}(t \sqrt{j 2 \pi f \mu_0 \sigma})\right)^{-1}
\end{equation}

\noindent where $t$ is the thickness of the superconductor, $\mu_0$ the permittivity of the material, and $f$ the frequency at which the impedance is evaluated. The complex conductivity, $\sigma = \sigma_1 + j\sigma_2$, of the superconductor under a given set of loading conditions is given by:\cite{gao2008}

\begin{equation}
     \sigma_1 = \sigma_N \frac{2 \Delta_0}{h f} \frac{n_{qp}}{N_0 \sqrt{2 \pi k_B T \Delta_0}} K_0(\xi)
 \end{equation}

\noindent and
\begin{equation}
 \sigma_2 = \sigma_N \frac{\pi \Delta_0}{hf} \left[1 - \frac{n_{qp}}{2N_0\Delta_0} \left(1 + \sqrt{\frac{2\Delta_0}{\pi k_B T}}e^{-\xi} I_0(\xi)\right)\right]
\end{equation}

\noindent where $\sigma_N$ is the normal-state conductivity for the sample, $I_0$ and $K_0$ are modified zeroth-order Bessel functions of the first and second kind respectively, $\xi = \frac{hf}{2k_B T}$, $\Delta_0$ is the zero-temperature gap energy, $h$ is the Planck constant, $N_0$ is the single spin density of states for the sample material (here, aluminum), and $k_B$ is the Boltzmann constant. The quasiparticle density, $n_{qp}$, is determined by the thermal and optical load on the detector, and we assume that it is unaffected by the readout current.

With $l$ and $w$ as the length and width of the inductor, we obtain values for the resistance and zero-current kinetic inductance as

\begin{equation}
    R = \mathrm{Re}(Z_s) \frac{l}{w} \quad \text{and} \quad L_k(0) = \frac{\mathrm{Im}(Z_s)}{2 \pi f} \frac{l}{w}
\end{equation}

\noindent and we assume that, by design, the other component values in the circuit ($C$, $L_g$, $C_c$) do not vary with absorbed energy of any kind. 

The detector's resonant frequency is the frequency at which $Z_{res}$ is minimized and its imaginary component is zero.
This frequency can generally be solved analytically as a function of all four lumped elements in the resonant circuit, and follows the trend $f_r \sim \sqrt{L (C + C_c)}^{-1}$.

\paragraph{Current-dependent inductance}

This work seeks primarily to demonstrate the use of readout current to manipulate the resonant frequency of a kinetic inductance detector. For simplicity, we consider only the reactive component of the resonator’s response to readout current. The validity of neglecting the dissipative response is touched upon in Section \ref{sec:amp_to_freq_sweep_mapping}. The resonator’s reactive response arises from the well-documented $I^2$ nonlinearity, which describes a current-dependent kinetic inductance of the form:

\begin{equation}\label{eq:Lk_nonlinearity}
L_k(I_{L}) \simeq L_k(0) \left( 1+\frac{|I_{L}|^2}{I_*^2} \right)
\end{equation}

\noindent where $I_{L}$ is the readout current flowing through the inductor, and $I_*$ sets the scale of the effect.\cite{pippard1950, zmuidzinas2012}

For a readout tone of amplitude $V_{in}$ at a given frequency $f_{drive}$ near a resonance, Eqs. \ref{eq:Ires} and \ref{eq:I_L} are now implicit expressions for the current flowing through (and therefore the impedance of) the resonator. They can be solved by beginning with the zero-current value for $Z_{res}(f_{drive})$ and iteratively evaluating the expression for $I_{L}$, using the value obtained after each iteration to update $L_k(I_{L})$. Proceeding in this manner, the system rapidly converges to stable values, representing the equilibrated impedance and current of the resonator for the given input. Once this is obtained, we compute the transfer function of the resonator as usual using Eq. \ref{eq:transfer_function}, with $Z_{res}$ now including the modification due to the current.

\begin{figure}[htbp]
\centering
\includegraphics[width=\linewidth]{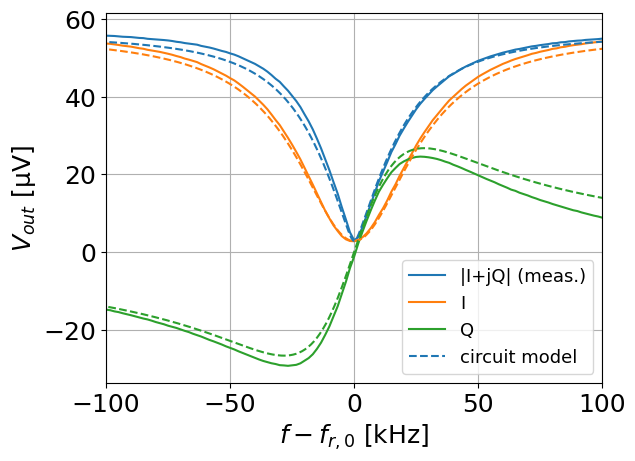}
\caption{Comparison of measured (at the input to the digitizer) complex output voltage for an example resonance (solid lines) versus the calculated output voltage (dashed lines) obtained using a circuit model based on this resonance. The model contains no stray impedances, resulting in an imperfect but sufficient agreement between the two. For consistency, this example resonance and circuit model are used in all figures throughout this work. The values of the circuit components, as well as the other parameters used for the resonator model (thermal and optical loading, inductor geometry, etc), may be found in Appendix \ref{app:model_params}. The I and Q voltages are by definition orthogonal to each other, but the global orientation of the IQ plane is arbitrary. Here the data has been rotated about the origin so that the derivative of Q is maximized at the resonant frequency, for the purposes of consistent visualization. These complex voltages can be related to the phase within the resonance bandwidth by fitting a circle to the data on the IQ plane and shifting the data so that the resonance circle is centred on the origin.}
\label{fig:circuit_model_vs_meas}
\end{figure}

We fit Eq. \ref{eq:transfer_function} to a measurement of a resonance’s response at low readout amplitude across its bandwidth to extract the circuit component values which cannot be obtained from the complex conductivity and inductor geometry. Combined with measurements of the gains and attenuations of the rest of the analog electronic system, we may then compute various aspects of the resonator’s response and transfer function, under the given loading and readout frequency drive conditions. Fig. \ref{fig:circuit_model_vs_meas} shows a comparison between the calculated and measured transfer function for an example resonance. The circuit model only considers the elements featured in Fig. \ref{fig:circuit_model}, and does not capture the effects of stray impedances which may be present in the physical system. As seen in Fig. \ref{fig:circuit_model_vs_meas}, this results in an agreement between the two that will be sufficient in this context as a tool in understanding the overall behaviour of the circuit.

A value for the scaling current, $I_*$, may be estimated by comparing the measured and predicted frequency shifts of a resonator in response to injected readout current. This method suffers from a challenge in separating the value of $I_*$ from the amplitude of the readout current reaching the resonator, which is difficult to measure precisely due to the many temperature- and geometry-dependent impedances in the system. Nonetheless, the resultant circuit model reproduces the observed behaviour with sufficient accuracy to usefully predict and examine many aspects of the readout-resonator interaction. For the resonators studied in this work, we find that values of $I_*$ in the range 0.5 - 1 mA closely reproduce the measured behaviour.

\subsection{Measurement setup}\label{sec:measurement}
A variety of lumped-element microwave kinetic inductance detectors were studied in the development of this work. \cite{dibert2022,dibert2023,barry2022}
The devices have aluminum inductors and niobium capacitors (both interdigitated and parallel plate). They are designed to be either direct absorbing, or to be part of an on-chip filterbank spectrometer. Their designs, fabrication procedures, and resonator properties vary widely, but none are thermally isolated from the wafer substrate. These devices were created as prototypes for future experiments on the South Pole Telescope\cite{anderson2022,karkare2021}, and were used here for the purposes of investigating the interaction of the readout current with the resonators. For consistency, all measurements shown in this work were made using a single resonator, arbitrarily selected from a prototype wafer containing 12 resonances of the style described in Ref. \onlinecite{dibert2022}. This resonator is representative of the others on this wafer, and the results and analysis in this work were found to be generally applicable to the other device types tested.

Measurements are made using RF-ICE, a multi-tone readout platform using polyphase filterbank synthesis and demodulation to operate up to 1024 sinusoidal carrier tones per readout line.\cite{bandura2016,rouble2022} The resonators are cryogenically cooled to between 25-200mK using a BlueFors LD250S dilution refrigerator. The resonators are mounted inside a dark aluminum sample box. The carrier comb is transmitted via coaxial cable from the synthesizer through approximately 50 dB of attenuation before the resonators, amplified by a cryogenic low-noise amplifier\cite{cryoelec} (LNA) after leaving the resonators, with additional stages of room-temperature amplification added as needed. A diagram of the measurement setup may be found in Fig. \ref{fig:circuit_model}a.

\section{Response to readout drive}\label{sec:readout_drive}

This work aims to develop a practical understanding of the response of a KID when driven with readout current, to enable the exertion of intentional control over the detector's resonant frequency using a fixed-frequency readout drive tone.
To achieve this, we first need to parametrize how the resonant frequency varies as a function of the applied drive amplitude and the frequency at which it is applied. Understanding these relations will allow us to determine the state of the KID as the readout current adjusts its resonant frequency, for the purposes of bringing it to a new operational bias point.

We reconstruct the resonance shape under a strong readout drive using multiple complementary methods. Given that the readout tone is providing the drive current with which we intend to alter the resonator's kinetic inductance, it is clear that sweeping that same readout tone in frequency across the resonant bandwidth will have marked effects on the system. The motion of the tone will be combined with the motion of the resonance in response to the tone, producing a complicated picture of a distorted resonance shape. This is generally referred to as resonance bifurcation.

The dynamic nature of this type of measurement makes it challenging to discern the complete transfer function of the resonance.
To disentangle the relevant effects, we instead place a tone at a fixed frequency, and vary its amplitude. This has two advantages: with the readout tone frequency now kept constant, only the motion of the resonance will be observed. Secondarily, this measurement technique mimics the intended control architecture, where we intend to operate and manipulate the resonant frequency of the detector using a fixed-frequency drive.

In the RF-ICE system, the programming of each carrier tone in the readout comb is fully independent in amplitude and frequency, and can be dynamically altered with negligible (sub-ms) latency. This allows a high degree of flexibility in the design of measurement algorithms.
As we vary the fixed-frequency drive tone amplitude, we monitor the state of the resonance in one of two ways.
One method uses multifrequency snapshot measurements as introduced in Ref. \onlinecite{rouble2024}, which employ additional small-amplitude readout tones to instantaneously measure the driven resonator's transfer function at an arbitrary number of frequency points across the bandwidth. The auxilliary tones are kept small enough that the power they deposit on the resonance is negligible compared to that deposited by the primary drive tone. This technique is straightforward to achieve with a multitone readout system, but an alternative measurement not requiring such a system could take the form of a single small secondary tone (such as from a VNA or standard homodyne system) being swept sequentially in frequency while the drive tone (contributed by a separate function generator) remains fixed.

The multifrequency measurements provide an easy visualization of the resonance's location under a given set of readout drive conditions. The use of this technique is complicated, however, by the onset of kinetic inductance parametric amplification, for which the $I^2$ nonlinearity is also the mechanism.
When the readout drive tone is within the resonance bandwidth and at sufficient amplitude, the small secondary tones used to probe the resonance experience parametric gain, as noted in this context in Fig. 3 of Ref. \onlinecite{rouble2024}. Once this amplification becomes dominant, the secondary tones used in the multifrequency measurement no longer probe the resonator's transfer function as a direct function of its impedance. The drive tone itself is not subject to the parametric gain, and (to a good approximation at moderate input powers) remains directly sensitive to the impedance of the resonator. Therefore, when the drive conditions are such that the parametric amplification is non-negligible, we revert to the second measurement option: relying on measurements from the drive tone itself in interpreting the results of the drive amplitude sweep.

In a drive amplitude sweep, as the drive tone amplitude is varied, the response of the resonator results in a changing impedance at the drive tone frequency. Measuring this impedance, and comparing it with the impedance measured as a function of frequency obtained through a readout tone frequency sweep across the same resonance at low amplitude, we can infer the location of the resonance at each point in the amplitude sweep. Sweeping the drive amplitude from low to high or high to low over a sufficient range of amplitudes, the motion of the resonance past the tone is such that the measurement traverses the entire bandwidth. Placing the drive tone at a range of offset frequencies from the unperturbed resonant frequency results in different amounts of readout current flowing through the resonance, for the same range of applied drive voltages. This is analogous to performing frequency sweeps across the resonance at a range of readout amplitudes. As seen in frequency sweeps, there is a region of this parameter space in which the response of the resonator is single-valued: an ascending sweep produces the same result as a descending sweep. Outside of this region, the resonator exhibits bistability, with a strong hysteresis on the direction of the sweep.

By combining these measurement techniques, we examine the resonator's response within these two regimes, and explore the behaviour of the resonant frequency as a function of the frequency and amplitude of the applied readout drive.

\subsection{Mapping between sweeps in amplitude and frequency}\label{sec:amp_to_freq_sweep_mapping} 

\begin{figure}[htbp]
    \centering
    \includegraphics[width=\linewidth]{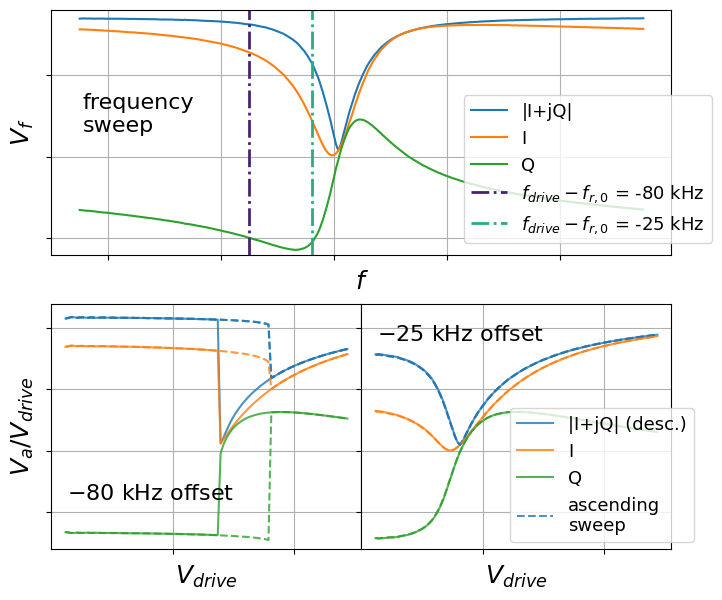}
    \caption{Comparison of constant-amplitude frequency sweep and fixed-frequency drive amplitude sweep measurements, for the same resonator. Top: low-amplitude sequential frequency sweep across the resonator, showing measured complex voltages, $V_f$ (magnitude blue, I orange, Q green), as a function of readout tone frequency, $f$. Vertical dot-dashed lines indicate the drive tone frequencies for the two amplitude sweep measurements, which are shown below. Lower panels: (left) amplitude sweep measurement with drive tone at -80 kHz offset from relaxed resonant frequency, $f_{r,0}$; (right) amplitude sweep measurement with drive tone at -25 kHz offset from $f_{r,0}$. Each measured output voltage, $V_a$, is normalized by the input voltage, $V_{drive}$, at which it was obtained. When the drive tone is initialized near to the relaxed resonant frequency as in the -25 kHz case, the system remains in the single-valued limit where the upwards and downwards amplitude sweeps (here shown overlaid) are identical. When it is initialized further from $f_{r,0}$, a directional hysteresis is evident.
In all panels, dashed lines indicate an ascending sweep direction (either in frequency or amplitude), and solid lines indicate a descending sweep. The two directions are only distinguishable in the -80 kHz offset amplitude sweep.}
    \label{fig:sweep_context}
\end{figure}

When characterizing the resonance transfer function as a function of frequency, it is customary to sweep a tone of fixed amplitude in frequency across the resonance's bandwidth. However, if we are instead interested in the response of the resonator to readout current, the motion of the probe tone complicates the picture, as the sequential frequency sweep combines the motion of the tone with the response of the resonator, while recording only one frequency at a time. We therefore take a less conventional approach to this problem, and use a readout tone which is fixed in frequency, but swept in amplitude. Fig. \ref{fig:sweep_context} compares these methods applied to the same resonator. The frequency sweep measurement (upper panel) is performed with a very low amplitude readout tone so as to not perturb the resonance. The two amplitude sweeps (lower panels) span the same range of amplitudes, but place the readout tone at different locations in frequency space (indicated by vertical lines superimposed on the top panel) relative to the starting location of the resonance. As can be seen in the lower panels of Fig. \ref{fig:sweep_context}, this leads to markedly different effects on the resonance.

According to Eq. \ref{eq:Lk_nonlinearity}, an increase in the current through the resonance will increase the kinetic inductance of the resonator and decrease its resonant frequency. If the readout tone which is supplying the current (the drive tone) is located lower in frequency than the undriven (relaxed) resonant frequency (i.e., $f_{drive} < f_{r,0}$), there is a positive feedback mechanism between the increase in drive amplitude and the resonant frequency, as the impedance at the drive frequency decreases as the resonance approaches the tone, which in turn increases the current through the resonance when driven with a fixed-amplitude drive voltage. This will enhance the impact of a drive amplitude increase, and for large drive amplitudes, results in the same runaway feedback effect and strong directional hysteresis that causes the canonical bifurcation behaviour seen in frequency sweep measurements. \cite{Swenson2013}
Conversely, for a drive tone placed above the resonant frequency ($f_{drive} > f_{r,0}$), negative feedback results from the increasing resonator impedance at the drive frequency as the resonance shifts to lower frequency, further from the tone.

When the drive tone is within the resonance bandwidth, parametric amplification is commonly observed for moderate and large drive amplitudes. This prevents the use of the multifrequency imaging mode to directly capture the resonance's response to a drive amplitude sweep in these cases. Instead, we can infer the location of the resonance using the measured return voltages $V_a(V_{drive})$ of the drive channel, and the voltages $V_f(f)$ gathered during a low-amplitude frequency sweep across the same resonator. Because both are a measure of the total circuit impedance, recorded complex voltages from the drive channel at each applied drive amplitude can be related to the complex voltages measured as a function of frequency during the frequency sweep, when both have been normalized by the drive amplitude at which they were obtained. By minimizing the Euclidean distance between each point in the amplitude sweep and the reference frequency sweep values, we can estimate the frequency distance between the drive tone and the resonant frequency. Said differently, for each amplitude sweep value $V_a(V_{drive})/V_{drive}$, we want to find the frequency $f$ that minimizes the expression

\begin{equation}\label{eq:sweep_mapping}
    \bigg| \frac{V_f(f)}{V_{b}} - \frac{V_a(V_{drive})}{V_{drive}} \bigg| 
\end{equation}

\noindent where we refer to the constant input amplitude of the readout tone used in the frequency sweep as $V_b$, for clarity in distinguishing it from the variable drive amplitude used in the amplitude sweep, which we call $V_{drive}$.

\begin{figure}[htbp]
    \centering
    \includegraphics[width=\linewidth]{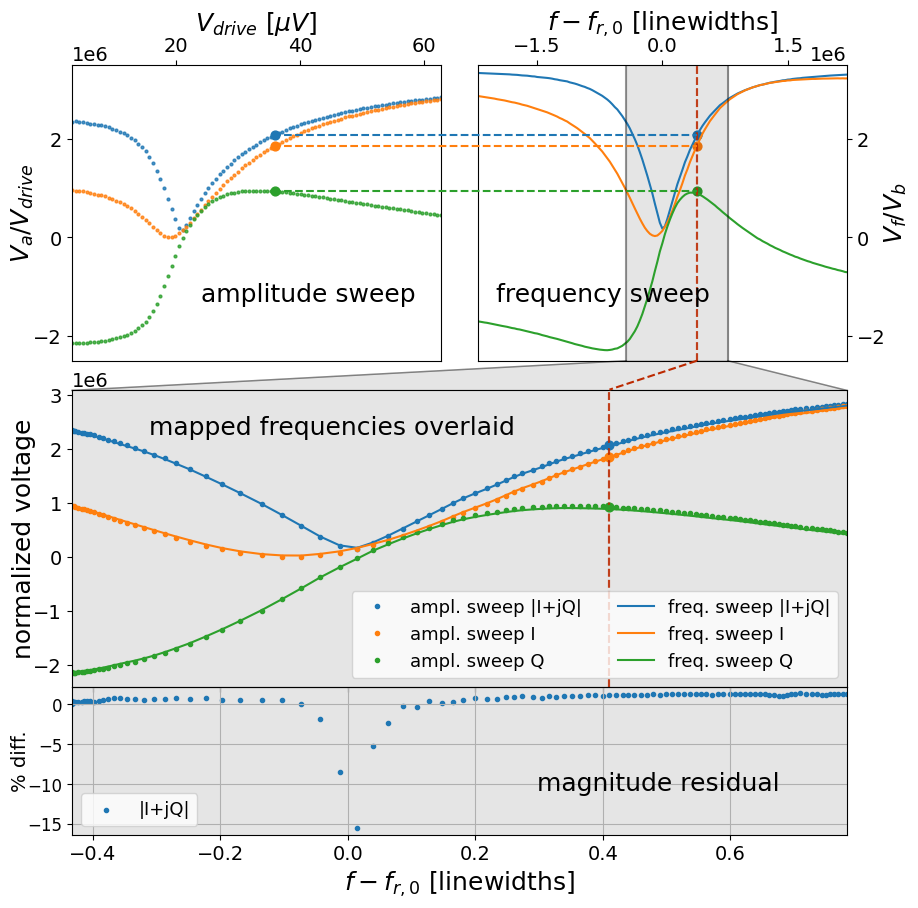}
    \caption{Determining the corresponding frequency within the resonance bandwidth for each measured voltage value of an amplitude sweep. Complex voltage values measured during the amplitude sweep are mapped to those obtained during a frequency sweep at low amplitude, by minimizing Eq. \ref{eq:sweep_mapping}. Overlaying the measured values from the amplitude sweep at their corresponding frequency locations within the frequency sweep (central panel; dots superimposed on solid lines), we see that the shape of the resonance remains largely unchanged, with the exception of the region $|f - f_{r}| \lesssim 0.1$ linewidths, wherein the depth of the resonance is seen to increase relative to its undriven shape by up to roughly 15\%. We therefore consider the effect of the readout current on the resonance to be primarily reactive.}
    \label{fig:mapping_ampsweep_to_freq}
\end{figure}

Fig. \ref{fig:mapping_ampsweep_to_freq} illustrates this procedure for the -25 kHz offset amplitude sweep measurement shown in the bottom right panel of Fig. \ref{fig:sweep_context}. Each measured complex voltage $V_a(V_{drive})/V_{drive}$ (Fig. \ref{fig:mapping_ampsweep_to_freq}, top left) is mapped to a measured complex value $V_f(f)/V_{b}$ from the frequency sweep (Fig. \ref{fig:mapping_ampsweep_to_freq}, top right), by minimizing Eq. \ref{eq:sweep_mapping}. In the lower panel, the amplitude sweep values (dots) are plotted at their mapped frequencies, $V_a(f)/V_{drive}$, and overlaid with the frequency sweep data (solid lines). The magnitude residual is included below to evaluate the goodness of the mapping. Once the corresponding frequency $f$ for each $V_a$ has been identified, the resonant frequency, $f_r$, at this drive amplitude may be computed as: 

\begin{equation}\label{eq:fr_mapping}
    f_r \big |_{V_{drive}} = f_{drive} - (f - f_{r,0}) \quad \text{.}
\end{equation}

\noindent Overlaying the measured values from the frequency sweep with those obtained from the amplitude sweep, at the extracted frequency location within the resonance bandwidth (main panel of Fig. \ref{fig:mapping_ampsweep_to_freq}) shows that, aside from an approximately 15\% increase in resonance depth when the resonant frequency is very close to the drive frequency, there is very little divergence between the two. This indicates that, over most of its bandwidth, although the resonator's impedance is changing as its inductance is modulated by the current, the overall shape of the transfer function remains largely unaffected, and the effect of the readout current on the resonator in this manner is primarily reactive. 
For simplicity in understanding the effects of readout current on resonant frequency within the scope of this work, we ignore the dissipative effects for the remainder of this work.

\subsection{Resonator response in the single-valued regime}\label{sec:single_valued_regime}

When $V_{drive}$ remains sufficiently small throughout an amplitude sweep, the response of the resonator is identical through both an increasing and decreasing sweep. In this single-valued regime, for a fixed-frequency drive tone with a given starting voltage amplitude, \textbf{the amplitude of the current through the resonator is entirely determined by the location of the drive tone relative to the relaxed resonant frequency.} The single-valued regime is therefore defined by a region in frequency space near to or above the relaxed resonance.

\begin{figure}[htbp]
\centering
\includegraphics[width=\linewidth]{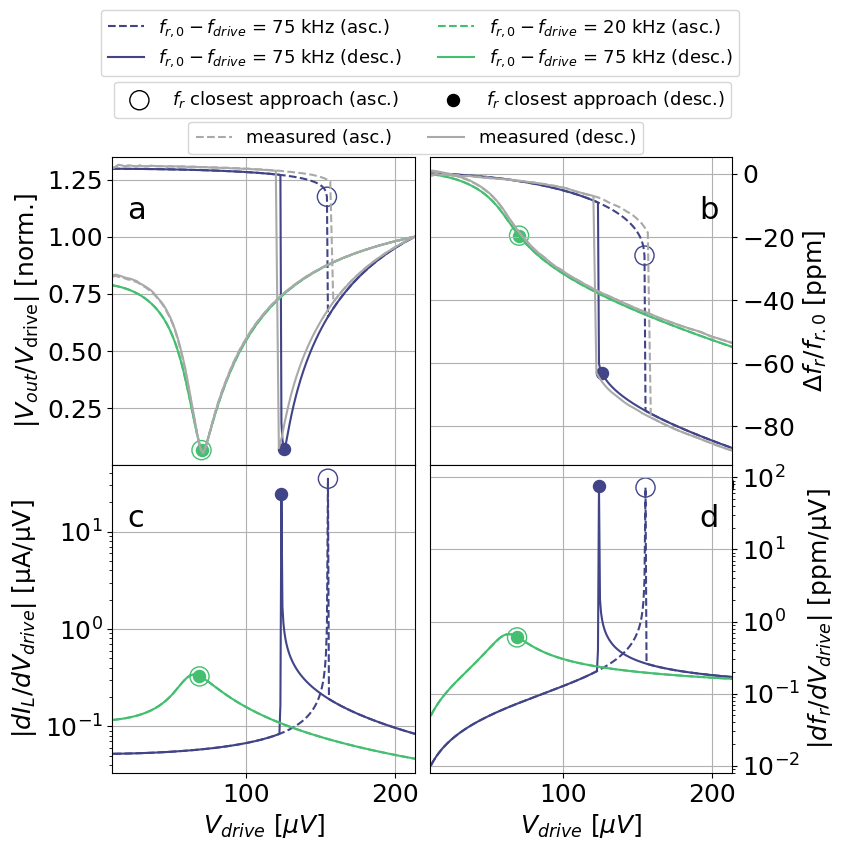}
\caption{Comparison and analysis of amplitude sweeps in the single-valued (green; starting offset -25 kHz) and bistable (blue; starting offset -80 kHz) regimes. Solid lines indicate a descending sweep direction, while dashed lines indicate an ascending sweep. Dots indicate the last stable point in each sweep before the resonant frequency passes the drive tone, with an open circle indicating this point on the ascending sweep and a closed circle for the descending sweep. In the single-valued regime, these points are the same in both directions, and are at the driven resonant frequency itself. In the -80 kHz sweep, the resonant frequency never reaches the drive tone, as runaway feedback causes the resonance to jump past the tone in each direction. Measured values are shown in grey, while all coloured traces are obtained by solving the circuit model with component values obtained by a fit to the measured resonance shape, as shown in Fig. \ref{fig:circuit_model_vs_meas}. (a) Output voltage at the drive tone frequency, normalized by input drive voltage. (b) Fractional change in driven resonant frequency versus relaxed resonant frequency. Here, grey traces are obtained by mapping the measured voltages to the unperturbed resonance shape (Eqs. \ref{eq:sweep_mapping} and \ref{eq:fr_mapping}).
(c) Rate of change of current through the inductor, with respect to change in the applied drive voltage. (d) Rate of change in driven resonant frequency, with respect to applied drive voltage. For both derivatives, in the -25 kHz offset sweep (the single-valued regime), the maximum occurs at the resonant frequency. For the -80 kHz offset sweep, the derivatives go to infinity before the resonant frequency reaches the drive tone, and the resonance jumps past the tone.}
\label{fig:amp_sweep_analysis}
\end{figure}

The green traces in Fig. \ref{fig:amp_sweep_analysis} examine the amplitude sweep measurement depicted in the lower right panel of Fig. \ref{fig:sweep_context}, which shows the resonator's single-valued response to a drive tone placed 25 kHz below the relaxed resonant frequency. From the measurement (overlaid as grey traces in Fig. \ref{fig:amp_sweep_analysis}), we obtain the output voltage and the change in resonant frequency (Fig. \ref{fig:amp_sweep_analysis}a and b, respectively). Using the circuit model, we can compute the associated evolution of the rates of change in the inductor current (Fig. \ref{fig:amp_sweep_analysis}c) and the resonant frequency (Fig. \ref{fig:amp_sweep_analysis}d) with respect to the applied drive amplitude.
As the drive amplitude is increased, the resonance moves toward the tone, reaching a new stable equilibrium resonant frequency at each input amplitude value, until eventually the resonant frequency moves smoothly past the drive tone and continues on to lower frequency. When the resonant frequency is above the drive tone, the impedance at the drive frequency decreases as the resonance moves towards the drive tone. Correspondingly, the rate of change in the current through the inductor increases with increasing proximity to the drive tone, as does the rate of change in the resonant frequency. This is a positive feedback regime.

When the resonant frequency is below the drive tone, the system enters the negative feedback regime: an increase in the drive amplitude results in an increase in the frequency distance between the resonance and the drive tone, which increases the impedance at the drive frequency and reduces the current flowing through the resonant branch (and thereby through the inductor). Although the impedance at the drive frequency quickly becomes large, because the applied drive voltage amplitude is steadily increased, the current through the resonance continues to increase as well, due to the increase in applied drive voltage. In this regime, the rate of change in the resonant frequency approaches a constant value, as can be seen at the higher drive amplitudes in Fig. \ref{fig:amp_sweep_analysis}d.

\begin{figure}[htbp]
\centering
\includegraphics[width=\linewidth]{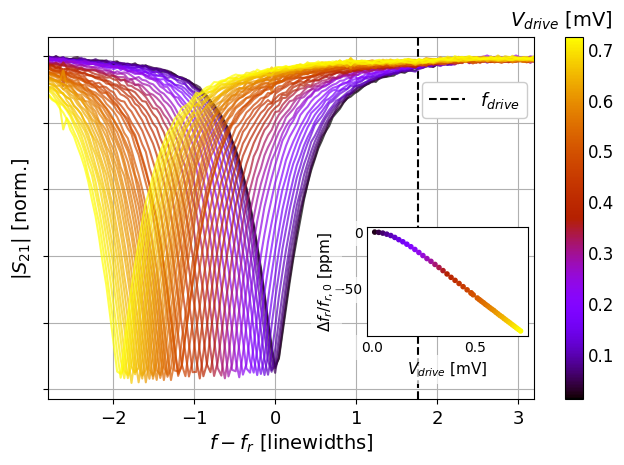}
\caption{\label{fig:driving_left_meas} Amplitude sweep for a drive tone at fixed frequency 100 kHz (approximately 2 linewidths; position indicated by dashed black line) above the relaxed resonant frequency. $|S_{21}|$ of the resonance is measured at each applied drive amplitude using multifrequency snapshot measurements (coloured traces). As the drive tone amplitude is increased, the resonance shifts to lower frequency, at a rate which is approximately linear in the applied amplitude.}
\end{figure}

A more striking demonstration of this effect is shown in Fig. \ref{fig:driving_left_meas}, where an amplitude sweep is performed with the drive tone initialized 100 kHz above the relaxed resonant frequency. Here we use multifrequency measurements as described in Section \ref{sec:measurement} (coloured traces) to directly capture the resonance transfer function as the drive amplitude is increased, and observe the approximately linear trend in the resonant frequency shift as a function of drive amplitude.

Proceeding in this manner, the resonance can be driven to a substantial frequency offset. The distance over which it may be driven is constrained by the linearity threshold of the LNA, which limits the allowed drive amplitude, and by current leakage through other resonators in the array. The nearest frequency neighbour to the resonance shown in Fig. \ref{fig:driving_left_meas} is roughly 2 MHz above its relaxed resonant frequency. When the target resonance has been driven away from the drive tone, a non-negligible fraction of the drive current will begin to flow through this neighbouring resonance. This increases its kinetic inductance, decreases its resonant frequency, and will eventually draw it in toward the drive tone, disrupting the further driving of the target resonance. In a more densely packed array, the nearest frequency neighbours may be expected to be closer. This drive tone leakage limits the extent of the amplitude that may be applied, unless the target is sufficiently isolated from its neighbours.

We now consider the utility of this technique as an operational mode. Typical readout of kinetic inductance detectors uses a single readout tone to monitor the location of the resonance in frequency space as it is modulated by absorbed light or thermal energy. The sensitivity of the readout tone to changes in the resonance falls as a function of the separation of the readout tone from the resonant frequency. Therefore, generally speaking, the further the resonance is driven from the drive tone, the less sensitive the drive tone is to changes in loading on the resonance. A second tone could be used to monitor the resonance in its new driven location, although this comes at the cost of increasing readout complexity.

However, if the drive tone is located below rather than above the relaxed resonant frequency, the readout amplitude may be sufficiently increased to bring the driven resonant frequency to align with the tone. At this point, the detector may be read out as usual, using a readout tone that is fixed in both amplitude and frequency. To the limit of not exceeding the linear dynamic range of the LNA, the resonant frequency of an arbitrary number of detectors in the array may be tuned in this fashion. Each resonant frequency may therefore be brought to the readout tone, rather than the readout needing to follow the resonance. This allows the use of integer-multiple frequency scheduling for intermodulation distortion mitigation,\cite{rouble2024} without sacrificing sensitivity due to detuning between the readout and resonant frequencies. Further, the use of larger readout amplitudes increases the ratio of the signal above the LNA noise, relaxing hardware constraints.

\subsection{Resonator response in the bistable regime}\label{sec:bistable_regime}

In the single-valued regime, for a given drive tone frequency at a given amplitude, there is a unique solution to the resonator impedance and resonant frequency. 
Outside this region (blue traces in Fig. \ref{fig:amp_sweep_analysis}), for upwards amplitude sweeps of drive tones below $f_{r,0}$, runaway positive feedback prevents the resonant frequency from ever stably reaching the drive tone. Instead, the resonance jumps past the tone, to a new stable state at lower frequency. This is the same instability which is seen as bifurcation in frequency sweeps at large constant amplitude (Refs. \onlinecite{deVisser2010, Swenson2013}, among others). In the frequency sweep context, the two branches of the bifurcated resonance state can be accessed by sweeping either upwards or downwards in frequency. Analogously, in an amplitude sweep, the second resonator state can be accessed by decreasing the amplitude of an initially-large drive tone. Unlike the single-valued regime, this is a region in the \textit{offset frequency-drive amplitude} parameter space where there are two physical solutions for the impedance and resonant frequency of a resonator, with each accessible through a hysteresis of the drive amplitude.

\begin{figure}[htbp]
    \centering
    \includegraphics[width=\linewidth]{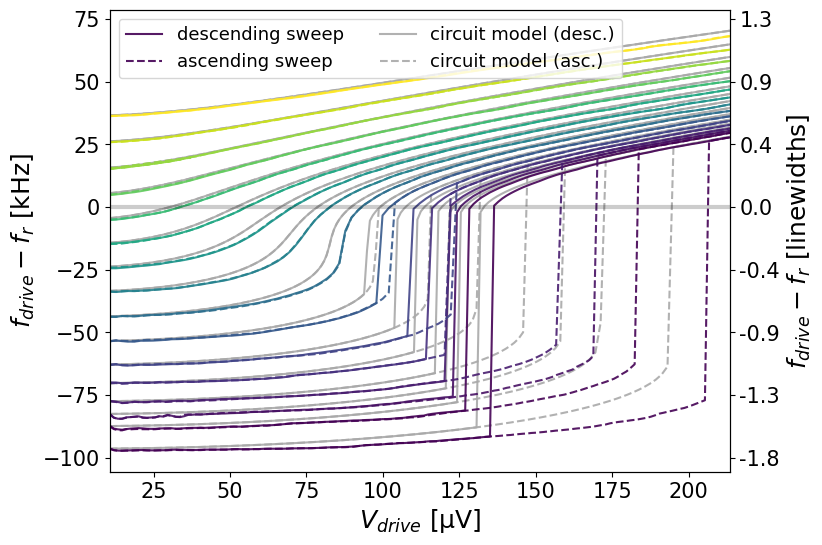}
    \caption{Resonant frequency as a function of readout drive amplitude (referred to the $V_{in}$ point in Fig. \ref{fig:circuit_model}) for amplitude sweeps taken over a range of starting drive frequency offsets from the relaxed resonant frequency. The coloured traces are measurements, with the colour of the trace corresponding to the starting frequency offset between the drive tone and the relaxed resonant frequency. Grey traces show the same quantity, computed using the circuit model based on this resonance. 
    For sweeps beginning with the drive tone below the relaxed resonant frequency, the resonance exhibits bistability for initial drive frequency offsets $f_{drive} - f_{r,0} \lesssim -50$ kHz. For sweeps beginning nearer to (small values of $|f_{drive} - f_{r,0}|$) or above the relaxed resonant frequency, the response is single-valued.
    Although on the ascending sweep, in the bistable region, the separation between the driven resonant frequency $f_r$ and the drive tone remains large, once the resonance has jumped to the second bistable state, $f_r$ may be approached from the other direction, by reducing $V_{drive}$.
    Any trace which has a stable point at or very near to the zero-crossing may be considered a viable operational bias point for the resonator. On the descending branch, such bias points may be achieved up to very large initial frequency offsets. However, due to the steepening of $df_r/dV_{drive}$ (Fig. \ref{fig:amp_sweep_analysis}d) at these points for larger initial offset frequencies, they are increasingly susceptible to being destabilized by electrical noise and/or quasiparticle fluctuations. Such susceptibility is the cause of the erratic jump points seen on the measured ascending sweeps, which here is particularly evident on sweeps in the range $-75$ kHz $\lesssim f_{drive} - f_{r,0} \lesssim -60$ kHz.}
    \label{fig:bistability_meas}
\end{figure}

This frequency-amplitude space is mapped out for an example resonator in Fig. \ref{fig:bistability_meas}. The drive tone is initialized at low amplitude at a range of offset frequencies from $f_{r,0}$. As its amplitude is increased over a range spanning approximately a factor of 20 in voltage, the measured voltage at $f_{drive}$ is recorded and used to determine the relative location of the driven resonant frequency, $f_r$, at each input amplitude via Eqs. \ref{eq:sweep_mapping} and \ref{eq:fr_mapping}. At each offset frequency, the drive amplitude is then swept downward from its maximum value, and $f_r$ again mapped as a function of input amplitude.

When the starting offset frequency is near to (small values of $|f_{drive} - f_{r,0}|$) or above the relaxed resonant frequency ($f_{drive} - f_{r,0} > 0$), the system is in the single-valued regime, where the upward and downward sweeps produce identical behaviour. For initial drive frequency offsets $f_{drive} - f_{r,0} \lesssim -50$ kHz, the resonance exhibits bistability. In this region, on the ascending sweep, the resonance jumps past the drive tone and the minimum attained separation between the two remains large. On the descending sweep, the resonance may be brought very close to the drive tone by slowly reducing the drive voltage. For this resonator, at the maximum frequency offset tested, the minimum attained separation via the descending branch was $\lesssim 1\%$ of the resonator linewidth. Any point along this branch which is sufficiently close to $f_{drive} - f_r = 0$ may be considered to be a viable operational bias point for the resonator, in terms of sensitivity to variation in the resonator transfer function due to changing optical load.

For starting frequency offsets beyond the onset of bistability, the rate of change of the driven resonant frequency with drive voltage, $df_r/dV_{drive}$ (Fig. \ref{fig:amp_sweep_analysis}d), becomes increasingly steep as $f_r$ approaches the drive tone. Accordingly, without the ability to quickly and actively modulate the drive amplitude to compensate for them, for larger starting offsets, ever smaller fluctuations in resonator impedance (e.g. due to fluctuations in load) may destabilize the system by driving it into the runaway feedback regime and ending up in the other bistable state. For more modest offsets, and especially within the single-valued regime, the resonator may be operated stably at its driven frequency over a wider range of changes in incident load. The requisite stability and dynamic range, and thereby the choice of offset frequency, will depend on the operational conditions and the expected signal loading variance.

\section{Conclusion}

We have examined the impact of readout current on the resonant frequency of kinetic inductance detectors, using a static-frequency readout drive tone.
The response of the resonators studied in this work is predominantly reactive in the range of drive amplitudes applied, resulting in modulation of the resonant frequency without significant change in the resonance shape, quality factor, or increased dissipation. We show that this reactive response can be understood using a simple lumped-element circuit model wherein the readout current modulates the kinetic inductance only. This model reproduces the observed changes in resonant frequency to a degree of accuracy sufficient to make it a useful predictive and interpretive tool in understanding the resonator–readout current interaction.

Varying the amplitude of a readout drive tone at fixed frequency near the resonant frequency of a target detector, we find that it is possible to alter the target’s resonant frequency by more than one linewidth. When the drive tone is located above the relaxed resonant frequency, increasing the drive amplitude pushes the resonance to lower frequency, further from the tone. When the tone is located below, the resonance is drawn toward it.

For small frequency offsets, the response is single-valued: there is a unique solution for the location of the resonant frequency for a given drive amplitude, and the resonant frequency can be altered so that it exactly matches the drive frequency. This is a stable operational bias point for the resonance, and the detector may be read out at its new driven resonant frequency using the fixed drive tone, in the same way as at its relaxed frequency.

At larger frequency offsets, runaway positive feedback leads to bistability: two stable states, accessible via hysteresis of the amplitude sweep. This is the same phenomenology referred to as resonance bifurcation when seen in constant-amplitude frequency sweep measurements. In this regime, in the ascending sweep, the driven resonant frequency is not accessible due to the runaway feedback, and there is therefore no stable operational point. On the descending branch, the driven resonant frequency can be relaxed by reducing the drive amplitude, so that it approaches the drive tone from below. In this study, the driven resonant frequency was brought to within a few percent of a linewidth over the range of initial offset frequencies studied. We postulate that this could be further extended, especially with the use of fast and active drive tone amplitude modulation to counteract instability due to thermal or optical loading fluctuations. Accessing the resonant frequency through this descending branch greatly extends the range over which the resonant frequency can be modified while maintaining a useful operational bias, as compared to the limited single-valued regime.

Relaxing the resonant frequency to the drive frequency in this way is analogous to the downward frequency sweep employed in Ref. \onlinecite{Swenson2013} to operate a resonator beyond its bifurcation power. The authors of that work noted improved device noise-equivalent power when operating in this regime. While we have not investigated device responsivity and noise in this work, our observations of resonant frequency behaviour as a function of readout current are consistent with theirs. The static frequency drive technique builds on their frequency-sweep analysis and allows the selection of the resonant frequency in advance, facilitating the tuning of operational bias points across an array. The ability to accurately model this behaviour using a simple lumped-element circuit model is well-suited to the analysis and understanding of these effects in the context of the rest of the readout system, and simplifies the development of novel readout–resonator control techniques.

We see great utility in being able to alter and control the resonant frequency of a KID in situ, without the need for specialty circuit designs or additional readout complexity. The ability to select and fine-tune resonant frequencies using a single traditional readout tone has implications for the mitigation of intermodulation distortion products and crosstalk, as well as the potential to relax fabrication complexity by correcting collisions between resonances without the need to physically trim capacitors.
Recent demonstrations have shown that it is possible to actively feed back on and control the kinetic inductance during operation to keep a KID’s resonant frequency fixed while the loading on it varies.\cite{rouble2024} Such a technique requires initializing the KID to some starting drive amplitude, which sets the dynamic range of the feedback controller. This work provides a framework for the understanding and manipulation of such a system, demonstrates that the starting drive amplitude which matches the resonant frequency to the drive tone is uniquely determined by the frequency separation of the drive tone from the relaxed resonant frequency, and that, via the descending branch of the amplitude sweep, a broad set of offset frequencies (and thereby wide dynamic range) are viable.

\section{Acknowledgements}
The authors wish to thank Peter Day for many illuminating discussions. 
The McGill authors acknowledge funding from the Natural Sciences and Engineering Research Council of Canada, the Canadian Foundation for Innovation, and the Canadian Institute for Advanced Research.
The detectors used in this work made use of the Pritzker Nanofabrication Facility of the Institute for Molecular Engineering at the University of Chicago, which receives support from Soft and Hybrid Nanotechnology Experimental (SHyNE) Resource (NSF ECCS-2025633), a node of the National Science Foundation’s National Nanotechnology Coordinated Infrastructure.

\nocite{*}
\bibliography{bibliography}

\providecommand{\noopsort}[1]{}\providecommand{\singleletter}[1]{#1}%
\begin{thebibliography}{23}%
\makeatletter
\providecommand \@ifxundefined [1]{%
 \@ifx{#1\undefined}
}%
\providecommand \@ifnum [1]{%
 \ifnum #1\expandafter \@firstoftwo
 \else \expandafter \@secondoftwo
 \fi
}%
\providecommand \@ifx [1]{%
 \ifx #1\expandafter \@firstoftwo
 \else \expandafter \@secondoftwo
 \fi
}%
\providecommand \natexlab [1]{#1}%
\providecommand \enquote  [1]{``#1''}%
\providecommand \bibnamefont  [1]{#1}%
\providecommand \bibfnamefont [1]{#1}%
\providecommand \citenamefont [1]{#1}%
\providecommand \href@noop [0]{\@secondoftwo}%
\providecommand \href [0]{\begingroup \@sanitize@url \@href}%
\providecommand \@href[1]{\@@startlink{#1}\@@href}%
\providecommand \@@href[1]{\endgroup#1\@@endlink}%
\providecommand \@sanitize@url [0]{\catcode `\\12\catcode `\$12\catcode
  `\&12\catcode `\#12\catcode `\^12\catcode `\_12\catcode `\%12\relax}%
\providecommand \@@startlink[1]{}%
\providecommand \@@endlink[0]{}%
\providecommand \url  [0]{\begingroup\@sanitize@url \@url }%
\providecommand \@url [1]{\endgroup\@href {#1}{\urlprefix }}%
\providecommand \urlprefix  [0]{URL }%
\providecommand \Eprint [0]{\href }%
\providecommand \doibase [0]{http://dx.doi.org/}%
\providecommand \selectlanguage [0]{\@gobble}%
\providecommand \bibinfo  [0]{\@secondoftwo}%
\providecommand \bibfield  [0]{\@secondoftwo}%
\providecommand \translation [1]{[#1]}%
\providecommand \BibitemOpen [0]{}%
\providecommand \bibitemStop [0]{}%
\providecommand \bibitemNoStop [0]{.\EOS\space}%
\providecommand \EOS [0]{\spacefactor3000\relax}%
\providecommand \BibitemShut  [1]{\csname bibitem#1\endcsname}%
\let\auto@bib@innerbib\@empty
\bibitem [{\citenamefont {Mazin}\ \emph {et~al.}(2002)\citenamefont {Mazin},
  \citenamefont {Day}, \citenamefont {Zmuidzinas},\ and\ \citenamefont
  {Leduc}}]{mazin2002}%
  \BibitemOpen
  \bibfield  {author} {\bibinfo {author} {\bibfnamefont {B.~A.}\ \bibnamefont
  {Mazin}}, \bibinfo {author} {\bibfnamefont {P.~K.}\ \bibnamefont {Day}},
  \bibinfo {author} {\bibfnamefont {J.}~\bibnamefont {Zmuidzinas}}, \ and\
  \bibinfo {author} {\bibfnamefont {H.~G.}\ \bibnamefont {Leduc}},\ }\bibfield
  {title} {\enquote {\bibinfo {title} {Multiplexable kinetic inductance
  detectors},}\ }\href {\doibase 10.1063/1.1457652} {\bibfield  {journal}
  {\bibinfo  {journal} {AIP Conference Proceedings}\ }\textbf {\bibinfo
  {volume} {605}},\ \bibinfo {pages} {309--312} (\bibinfo {year} {2002})},\
  \Eprint
  {http://arxiv.org/abs/https://pubs.aip.org/aip/acp/article-pdf/605/1/309/11803582/309\_1\_online.pdf}
  {https://pubs.aip.org/aip/acp/article-pdf/605/1/309/11803582/309\_1\_online.pdf}
  \BibitemShut {NoStop}%
\bibitem [{\citenamefont {Day}\ \emph {et~al.}(2003)\citenamefont {Day},
  \citenamefont {LeDuc}, \citenamefont {Mazin} \emph {et~al.}}]{day2003}%
  \BibitemOpen
  \bibfield  {author} {\bibinfo {author} {\bibfnamefont {P.}~\bibnamefont
  {Day}}, \bibinfo {author} {\bibfnamefont {H.}~\bibnamefont {LeDuc}}, \bibinfo
  {author} {\bibfnamefont {B.}~\bibnamefont {Mazin}},  \emph {et~al.},\
  }\bibfield  {title} {\enquote {\bibinfo {title} {A broadband superconducting
  detector suitable for use in large arrays},}\ }\href {\doibase
  10.1038/nature02037} {\bibfield  {journal} {\bibinfo  {journal} {Nature}\
  }\textbf {\bibinfo {volume} {425}},\ \bibinfo {pages} {817--821} (\bibinfo
  {year} {2003})}\BibitemShut {NoStop}%
\bibitem [{\citenamefont {Endo}\ \emph {et~al.}(2019)\citenamefont {Endo},
  \citenamefont {Karatsu}, \citenamefont {Tamura}, \citenamefont {Oshima},
  \citenamefont {Taniguchi}, \citenamefont {Takekoshi}, \citenamefont
  {Asayama}, \citenamefont {Bakx}, \citenamefont {Bosma}, \citenamefont
  {Bueno}, \citenamefont {Chin}, \citenamefont {Fujii}, \citenamefont {Fujita},
  \citenamefont {Huiting}, \citenamefont {Ikarashi}, \citenamefont {Ishida},
  \citenamefont {Ishii}, \citenamefont {Kawabe}, \citenamefont {Klapwijk},
  \citenamefont {Kohno}, \citenamefont {Kouchi}, \citenamefont {Llombart},
  \citenamefont {Maekawa}, \citenamefont {Murugesan}, \citenamefont
  {Nakatsubo}, \citenamefont {Naruse}, \citenamefont {Ohtawara}, \citenamefont
  {Pascual~Laguna}, \citenamefont {Suzuki}, \citenamefont {Suzuki},
  \citenamefont {Thoen}, \citenamefont {Tsukagoshi}, \citenamefont {Ueda},
  \citenamefont {de~Visser}, \citenamefont {van~der Werf}, \citenamefont
  {Yates}, \citenamefont {Yoshimura}, \citenamefont {Yurduseven},\ and\
  \citenamefont {Baselmans}}]{endo2019}%
  \BibitemOpen
  \bibfield  {author} {\bibinfo {author} {\bibfnamefont {A.}~\bibnamefont
  {Endo}}, \bibinfo {author} {\bibfnamefont {K.}~\bibnamefont {Karatsu}},
  \bibinfo {author} {\bibfnamefont {Y.}~\bibnamefont {Tamura}}, \bibinfo
  {author} {\bibfnamefont {T.}~\bibnamefont {Oshima}}, \bibinfo {author}
  {\bibfnamefont {A.}~\bibnamefont {Taniguchi}}, \bibinfo {author}
  {\bibfnamefont {T.}~\bibnamefont {Takekoshi}}, \bibinfo {author}
  {\bibfnamefont {S.}~\bibnamefont {Asayama}}, \bibinfo {author} {\bibfnamefont
  {T.~J. L.~C.}\ \bibnamefont {Bakx}}, \bibinfo {author} {\bibfnamefont
  {S.}~\bibnamefont {Bosma}}, \bibinfo {author} {\bibfnamefont
  {J.}~\bibnamefont {Bueno}}, \bibinfo {author} {\bibfnamefont {K.~W.}\
  \bibnamefont {Chin}}, \bibinfo {author} {\bibfnamefont {Y.}~\bibnamefont
  {Fujii}}, \bibinfo {author} {\bibfnamefont {K.}~\bibnamefont {Fujita}},
  \bibinfo {author} {\bibfnamefont {R.}~\bibnamefont {Huiting}}, \bibinfo
  {author} {\bibfnamefont {S.}~\bibnamefont {Ikarashi}}, \bibinfo {author}
  {\bibfnamefont {T.}~\bibnamefont {Ishida}}, \bibinfo {author} {\bibfnamefont
  {S.}~\bibnamefont {Ishii}}, \bibinfo {author} {\bibfnamefont
  {R.}~\bibnamefont {Kawabe}}, \bibinfo {author} {\bibfnamefont {T.~M.}\
  \bibnamefont {Klapwijk}}, \bibinfo {author} {\bibfnamefont {K.}~\bibnamefont
  {Kohno}}, \bibinfo {author} {\bibfnamefont {A.}~\bibnamefont {Kouchi}},
  \bibinfo {author} {\bibfnamefont {N.}~\bibnamefont {Llombart}}, \bibinfo
  {author} {\bibfnamefont {J.}~\bibnamefont {Maekawa}}, \bibinfo {author}
  {\bibfnamefont {V.}~\bibnamefont {Murugesan}}, \bibinfo {author}
  {\bibfnamefont {S.}~\bibnamefont {Nakatsubo}}, \bibinfo {author}
  {\bibfnamefont {M.}~\bibnamefont {Naruse}}, \bibinfo {author} {\bibfnamefont
  {K.}~\bibnamefont {Ohtawara}}, \bibinfo {author} {\bibfnamefont
  {A.}~\bibnamefont {Pascual~Laguna}}, \bibinfo {author} {\bibfnamefont
  {J.}~\bibnamefont {Suzuki}}, \bibinfo {author} {\bibfnamefont
  {K.}~\bibnamefont {Suzuki}}, \bibinfo {author} {\bibfnamefont {D.~J.}\
  \bibnamefont {Thoen}}, \bibinfo {author} {\bibfnamefont {T.}~\bibnamefont
  {Tsukagoshi}}, \bibinfo {author} {\bibfnamefont {T.}~\bibnamefont {Ueda}},
  \bibinfo {author} {\bibfnamefont {P.~J.}\ \bibnamefont {de~Visser}}, \bibinfo
  {author} {\bibfnamefont {P.~P.}\ \bibnamefont {van~der Werf}}, \bibinfo
  {author} {\bibfnamefont {S.~J.~C.}\ \bibnamefont {Yates}}, \bibinfo {author}
  {\bibfnamefont {Y.}~\bibnamefont {Yoshimura}}, \bibinfo {author}
  {\bibfnamefont {O.}~\bibnamefont {Yurduseven}}, \ and\ \bibinfo {author}
  {\bibfnamefont {J.~J.~A.}\ \bibnamefont {Baselmans}},\ }\bibfield  {title}
  {\enquote {\bibinfo {title} {{First light demonstration of the integrated
  superconducting spectrometer}},}\ }\href {\doibase 10.1038/s41550-019-0850-8}
  {\bibfield  {journal} {\bibinfo  {journal} {Nature Astronomy}\ }\textbf
  {\bibinfo {volume} {3}},\ \bibinfo {pages} {989--996} (\bibinfo {year}
  {2019})}\BibitemShut {NoStop}%
\bibitem [{\citenamefont {{Karkare}}\ \emph {et~al.}(2022)\citenamefont
  {{Karkare}}, \citenamefont {{Anderson}}, \citenamefont {{Barry}},
  \citenamefont {{Benson}}, \citenamefont {{Carlstrom}}, \citenamefont
  {{Cecil}}, \citenamefont {{Chang}}, \citenamefont {{Dobbs}}, \citenamefont
  {{Hollister}}, \citenamefont {{Keating}}, \citenamefont {{Marrone}},
  \citenamefont {{McMahon}}, \citenamefont {{Montgomery}}, \citenamefont
  {{Pan}}, \citenamefont {{Robson}}, \citenamefont {{Rouble}}, \citenamefont
  {{Shirokoff}},\ and\ \citenamefont {{Smecher}}}]{karkare2021}%
  \BibitemOpen
  \bibfield  {author} {\bibinfo {author} {\bibfnamefont {K.~S.}\ \bibnamefont
  {{Karkare}}}, \bibinfo {author} {\bibfnamefont {A.~J.}\ \bibnamefont
  {{Anderson}}}, \bibinfo {author} {\bibfnamefont {P.~S.}\ \bibnamefont
  {{Barry}}}, \bibinfo {author} {\bibfnamefont {B.~A.}\ \bibnamefont
  {{Benson}}}, \bibinfo {author} {\bibfnamefont {J.~E.}\ \bibnamefont
  {{Carlstrom}}}, \bibinfo {author} {\bibfnamefont {T.}~\bibnamefont
  {{Cecil}}}, \bibinfo {author} {\bibfnamefont {C.~L.}\ \bibnamefont
  {{Chang}}}, \bibinfo {author} {\bibfnamefont {M.~A.}\ \bibnamefont
  {{Dobbs}}}, \bibinfo {author} {\bibfnamefont {M.}~\bibnamefont
  {{Hollister}}}, \bibinfo {author} {\bibfnamefont {G.~K.}\ \bibnamefont
  {{Keating}}}, \bibinfo {author} {\bibfnamefont {D.~P.}\ \bibnamefont
  {{Marrone}}}, \bibinfo {author} {\bibfnamefont {J.}~\bibnamefont
  {{McMahon}}}, \bibinfo {author} {\bibfnamefont {J.}~\bibnamefont
  {{Montgomery}}}, \bibinfo {author} {\bibfnamefont {Z.}~\bibnamefont {{Pan}}},
  \bibinfo {author} {\bibfnamefont {G.}~\bibnamefont {{Robson}}}, \bibinfo
  {author} {\bibfnamefont {M.}~\bibnamefont {{Rouble}}}, \bibinfo {author}
  {\bibfnamefont {E.}~\bibnamefont {{Shirokoff}}}, \ and\ \bibinfo {author}
  {\bibfnamefont {G.}~\bibnamefont {{Smecher}}},\ }\bibfield  {title} {\enquote
  {\bibinfo {title} {{SPT-SLIM: A Line Intensity Mapping Pathfinder for the
  South Pole Telescope}},}\ }\href {\doibase 10.1007/s10909-022-02702-2}
  {\bibfield  {journal} {\bibinfo  {journal} {Journal of Low Temperature
  Physics}\ } (\bibinfo {year} {2022}),\
  10.1007/s10909-022-02702-2}\BibitemShut {NoStop}%
\bibitem [{\citenamefont {Vissers}\ \emph {et~al.}(2024)\citenamefont
  {Vissers}, \citenamefont {Wheeler}, \citenamefont {Austermann}, \citenamefont
  {Vaskuri}, \citenamefont {Hubmayr}, \citenamefont {Gao}, \citenamefont
  {Huber}, \citenamefont {Imrek},\ and\ \citenamefont {Ullom}}]{vissers2024}%
  \BibitemOpen
  \bibfield  {author} {\bibinfo {author} {\bibfnamefont {M.~R.}\ \bibnamefont
  {Vissers}}, \bibinfo {author} {\bibfnamefont {J.}~\bibnamefont {Wheeler}},
  \bibinfo {author} {\bibfnamefont {J.}~\bibnamefont {Austermann}}, \bibinfo
  {author} {\bibfnamefont {A.}~\bibnamefont {Vaskuri}}, \bibinfo {author}
  {\bibfnamefont {J.}~\bibnamefont {Hubmayr}}, \bibinfo {author} {\bibfnamefont
  {J.}~\bibnamefont {Gao}}, \bibinfo {author} {\bibfnamefont {Z.}~\bibnamefont
  {Huber}}, \bibinfo {author} {\bibfnamefont {J.}~\bibnamefont {Imrek}}, \ and\
  \bibinfo {author} {\bibfnamefont {J.}~\bibnamefont {Ullom}},\ }\bibfield
  {title} {\enquote {\bibinfo {title} {{Improving the yield of CCAT MKID arrays
  with post-measurement lithographic corrections}},}\ }in\ \href {\doibase
  10.1117/12.3020661} {\emph {\bibinfo {booktitle} {Millimeter, Submillimeter,
  and Far-Infrared Detectors and Instrumentation for Astronomy XII}}},\ Vol.\
  \bibinfo {volume} {PC13102},\ \bibinfo {editor} {edited by\ \bibinfo {editor}
  {\bibfnamefont {J.}~\bibnamefont {Zmuidzinas}}\ and\ \bibinfo {editor}
  {\bibfnamefont {J.-R.}\ \bibnamefont {Gao}}},\ \bibinfo {organization}
  {International Society for Optics and Photonics}\ (\bibinfo  {publisher}
  {SPIE},\ \bibinfo {year} {2024})\ p.\ \bibinfo {pages}
  {PC131020Z}\BibitemShut {NoStop}%
\bibitem [{\citenamefont {Vissers}\ \emph {et~al.}(2015)\citenamefont
  {Vissers}, \citenamefont {Hubmayr}, \citenamefont {Sandberg}, \citenamefont
  {Chaudhuri}, \citenamefont {Bockstiegel},\ and\ \citenamefont
  {Gao}}]{vissers2015}%
  \BibitemOpen
  \bibfield  {author} {\bibinfo {author} {\bibfnamefont {M.~R.}\ \bibnamefont
  {Vissers}}, \bibinfo {author} {\bibfnamefont {J.}~\bibnamefont {Hubmayr}},
  \bibinfo {author} {\bibfnamefont {M.}~\bibnamefont {Sandberg}}, \bibinfo
  {author} {\bibfnamefont {S.}~\bibnamefont {Chaudhuri}}, \bibinfo {author}
  {\bibfnamefont {C.}~\bibnamefont {Bockstiegel}}, \ and\ \bibinfo {author}
  {\bibfnamefont {J.}~\bibnamefont {Gao}},\ }\bibfield  {title} {\enquote
  {\bibinfo {title} {Frequency-tunable superconducting resonators via nonlinear
  kinetic inductance},}\ }\href {\doibase 10.1063/1.4927444} {\bibfield
  {journal} {\bibinfo  {journal} {Applied Physics Letters}\ }\textbf {\bibinfo
  {volume} {107}},\ \bibinfo {pages} {062601} (\bibinfo {year} {2015})},\
  \Eprint
  {http://arxiv.org/abs/https://pubs.aip.org/aip/apl/article-pdf/doi/10.1063/1.4927444/14466852/062601\_1\_online.pdf}
  {https://pubs.aip.org/aip/apl/article-pdf/doi/10.1063/1.4927444/14466852/062601\_1\_online.pdf}
  \BibitemShut {NoStop}%
\bibitem [{\citenamefont {de~Visser}, \citenamefont {Withington},\ and\
  \citenamefont {Goldie}(2010)}]{deVisser2010}%
  \BibitemOpen
  \bibfield  {author} {\bibinfo {author} {\bibfnamefont {P.~J.}\ \bibnamefont
  {de~Visser}}, \bibinfo {author} {\bibfnamefont {S.}~\bibnamefont
  {Withington}}, \ and\ \bibinfo {author} {\bibfnamefont {D.~J.}\ \bibnamefont
  {Goldie}},\ }\bibfield  {title} {\enquote {\bibinfo {title} {{Readout-power
  heating and hysteretic switching between thermal quasiparticle states in
  kinetic inductance detectors}},}\ }\href {\doibase 10.1063/1.3517152}
  {\bibfield  {journal} {\bibinfo  {journal} {Journal of Applied Physics}\
  }\textbf {\bibinfo {volume} {108}},\ \bibinfo {pages} {114504} (\bibinfo
  {year} {2010})}\BibitemShut {NoStop}%
\bibitem [{\citenamefont {Swenson}\ \emph {et~al.}(2013)\citenamefont
  {Swenson}, \citenamefont {Day}, \citenamefont {Eom}, \citenamefont {Leduc},
  \citenamefont {Llombart}, \citenamefont {McKenney}, \citenamefont
  {Noroozian},\ and\ \citenamefont {Zmuidzinas}}]{Swenson2013}%
  \BibitemOpen
  \bibfield  {author} {\bibinfo {author} {\bibfnamefont {L.~J.}\ \bibnamefont
  {Swenson}}, \bibinfo {author} {\bibfnamefont {P.~K.}\ \bibnamefont {Day}},
  \bibinfo {author} {\bibfnamefont {B.~H.}\ \bibnamefont {Eom}}, \bibinfo
  {author} {\bibfnamefont {H.~G.}\ \bibnamefont {Leduc}}, \bibinfo {author}
  {\bibfnamefont {N.}~\bibnamefont {Llombart}}, \bibinfo {author}
  {\bibfnamefont {C.~M.}\ \bibnamefont {McKenney}}, \bibinfo {author}
  {\bibfnamefont {O.}~\bibnamefont {Noroozian}}, \ and\ \bibinfo {author}
  {\bibfnamefont {J.}~\bibnamefont {Zmuidzinas}},\ }\bibfield  {title}
  {\enquote {\bibinfo {title} {Operation of a titanium nitride superconducting
  microresonator detector in the nonlinear regime},}\ }\href {\doibase
  10.1063/1.4794808} {\bibfield  {journal} {\bibinfo  {journal} {Journal of
  Applied Physics}\ }\textbf {\bibinfo {volume} {113}},\ \bibinfo {pages}
  {104501} (\bibinfo {year} {2013})}\BibitemShut {NoStop}%
\bibitem [{\citenamefont {Rouble}\ \emph {et~al.}(2024)\citenamefont {Rouble},
  \citenamefont {Smecher}, \citenamefont {Adamič}, \citenamefont {Anderson},
  \citenamefont {Barry}, \citenamefont {Dibert}, \citenamefont {Dobbs},
  \citenamefont {Fichman},\ and\ \citenamefont {Montgomery}}]{rouble2024}%
  \BibitemOpen
  \bibfield  {author} {\bibinfo {author} {\bibfnamefont {M.}~\bibnamefont
  {Rouble}}, \bibinfo {author} {\bibfnamefont {G.}~\bibnamefont {Smecher}},
  \bibinfo {author} {\bibfnamefont {M.}~\bibnamefont {Adamič}}, \bibinfo
  {author} {\bibfnamefont {A.}~\bibnamefont {Anderson}}, \bibinfo {author}
  {\bibfnamefont {P.~S.}\ \bibnamefont {Barry}}, \bibinfo {author}
  {\bibfnamefont {K.}~\bibnamefont {Dibert}}, \bibinfo {author} {\bibfnamefont
  {M.}~\bibnamefont {Dobbs}}, \bibinfo {author} {\bibfnamefont
  {K.}~\bibnamefont {Fichman}}, \ and\ \bibinfo {author} {\bibfnamefont
  {J.}~\bibnamefont {Montgomery}},\ }\bibfield  {title} {\enquote {\bibinfo
  {title} {A first demonstration of active feedback control and multifrequency
  imaging techniques for kinetic inductance detectors},}\ }in\ \href {\doibase
  10.1117/12.3019352} {\emph {\bibinfo {booktitle} {Millimeter, Submillimeter,
  and Far-Infrared Detectors and Instrumentation for Astronomy XII}}},\
  \bibinfo {series} {Proceedings of SPIE}, Vol.\ \bibinfo {volume} {13102},\
  \bibinfo {editor} {edited by\ \bibinfo {editor} {\bibfnamefont
  {J.}~\bibnamefont {Zmuidzinas}}\ and\ \bibinfo {editor} {\bibfnamefont
  {J.-R.}\ \bibnamefont {Gao}}}\ (\bibinfo {year} {2024})\ p.\ \bibinfo {pages}
  {131020Q}\BibitemShut {NoStop}%
\bibitem [{\citenamefont {Rouble}\ \emph {et~al.}(2022)\citenamefont {Rouble},
  \citenamefont {Smecher}, \citenamefont {Anderson}, \citenamefont {Barry},
  \citenamefont {Dibert}, \citenamefont {Dobbs}, \citenamefont {Karkare},\ and\
  \citenamefont {Montgomery}}]{rouble2022}%
  \BibitemOpen
  \bibfield  {author} {\bibinfo {author} {\bibfnamefont {M.}~\bibnamefont
  {Rouble}}, \bibinfo {author} {\bibfnamefont {G.}~\bibnamefont {Smecher}},
  \bibinfo {author} {\bibfnamefont {A.}~\bibnamefont {Anderson}}, \bibinfo
  {author} {\bibfnamefont {P.~S.}\ \bibnamefont {Barry}}, \bibinfo {author}
  {\bibfnamefont {K.}~\bibnamefont {Dibert}}, \bibinfo {author} {\bibfnamefont
  {M.}~\bibnamefont {Dobbs}}, \bibinfo {author} {\bibfnamefont {K.~S.}\
  \bibnamefont {Karkare}}, \ and\ \bibinfo {author} {\bibfnamefont
  {J.}~\bibnamefont {Montgomery}},\ }\bibfield  {title} {\enquote {\bibinfo
  {title} {{RF‐ICE: large‐scale gigahertz readout of
  frequency‐multiplexed microwave kinetic inductance detectors}},}\ }in\
  \href {\doibase 10.1117/12.2630286} {\emph {\bibinfo {booktitle} {Millimeter,
  Submillimeter, and Far‐Infrared Detectors and Instrumentation for Astronomy
  XI}}},\ \bibinfo {series} {Proceedings of SPIE}, Vol.\ \bibinfo {volume}
  {12190}\ (\bibinfo {year} {2022})\ p.\ \bibinfo {pages} {1219024}\BibitemShut
  {NoStop}%
\bibitem [{\citenamefont {Henkels}\ and\ \citenamefont
  {Kircher}(1977)}]{henkels1977}%
  \BibitemOpen
  \bibfield  {author} {\bibinfo {author} {\bibfnamefont {W.}~\bibnamefont
  {Henkels}}\ and\ \bibinfo {author} {\bibfnamefont {C.}~\bibnamefont
  {Kircher}},\ }\bibfield  {title} {\enquote {\bibinfo {title} {{Penetration
  depth measurements on type II superconducting films}},}\ }\href {\doibase
  10.1109/TMAG.1977.1059426} {\bibfield  {journal} {\bibinfo  {journal} {IEEE
  Transactions on Magnetics}\ }\textbf {\bibinfo {volume} {13}},\ \bibinfo
  {pages} {63--66} (\bibinfo {year} {1977})}\BibitemShut {NoStop}%
\bibitem [{\citenamefont {de~Visser}(2014)}]{deVisser2014}%
  \BibitemOpen
  \bibfield  {author} {\bibinfo {author} {\bibfnamefont {P.~J.}\ \bibnamefont
  {de~Visser}},\ }\emph {\bibinfo {title} {Quasiparticle dynamics in aluminium
  superconducting microwave resonators}},\ \href@noop {} {Ph.D. thesis},\
  \bibinfo  {school} {Technische Universiteit Delft} (\bibinfo {year} {2014}),\
  \bibinfo {note} {ph.D. thesis}\BibitemShut {NoStop}%
\bibitem [{\citenamefont {Gao}(2008)}]{gao2008}%
  \BibitemOpen
  \bibfield  {author} {\bibinfo {author} {\bibfnamefont {J.}~\bibnamefont
  {Gao}},\ }\emph {\bibinfo {title} {The Physics of Superconducting Microwave
  Resonators}},\ \href@noop {} {Ph.D. thesis},\ \bibinfo  {school} {California
  Institute of Technology} (\bibinfo {year} {2008}),\ \bibinfo {note} {ph.D.
  thesis}\BibitemShut {NoStop}%
\bibitem [{\citenamefont {Pippard}(1950)}]{pippard1950}%
  \BibitemOpen
  \bibfield  {author} {\bibinfo {author} {\bibfnamefont {A.~B.}\ \bibnamefont
  {Pippard}},\ }\bibfield  {title} {\enquote {\bibinfo {title} {Field variation
  of the superconducting penetration depth},}\ }\href@noop {} {\bibfield
  {journal} {\bibinfo  {journal} {Proceedings of the Royal Society of London.
  Series A, Mathematical and Physical Sciences}\ }\textbf {\bibinfo {volume}
  {203}},\ \bibinfo {pages} {210--223} (\bibinfo {year} {1950})}\BibitemShut
  {NoStop}%
\bibitem [{\citenamefont {Zmuidzinas}(2012)}]{zmuidzinas2012}%
  \BibitemOpen
  \bibfield  {author} {\bibinfo {author} {\bibfnamefont {J.}~\bibnamefont
  {Zmuidzinas}},\ }\bibfield  {title} {\enquote {\bibinfo {title}
  {Superconducting microresonators: Physics and applications},}\ }\href
  {\doibase https://doi.org/10.1146/annurev-conmatphys-020911-125022}
  {\bibfield  {journal} {\bibinfo  {journal} {Annual Review of Condensed Matter
  Physics}\ }\textbf {\bibinfo {volume} {3}},\ \bibinfo {pages} {169--214}
  (\bibinfo {year} {2012})}\BibitemShut {NoStop}%
\bibitem [{\citenamefont {Dibert}\ \emph {et~al.}(2022)\citenamefont {Dibert},
  \citenamefont {Barry}, \citenamefont {Pan},\ and\ \citenamefont
  {et~al.}}]{dibert2022}%
  \BibitemOpen
  \bibfield  {author} {\bibinfo {author} {\bibfnamefont {K.}~\bibnamefont
  {Dibert}}, \bibinfo {author} {\bibfnamefont {P.}~\bibnamefont {Barry}},
  \bibinfo {author} {\bibfnamefont {Z.}~\bibnamefont {Pan}}, \ and\ \bibinfo
  {author} {\bibnamefont {et~al.}},\ }\bibfield  {title} {\enquote {\bibinfo
  {title} {{Development of MKIDs for Measurement of the Cosmic Microwave
  Background with the South Pole Telescope}},}\ }\href {\doibase
  10.1007/s10909-022-02750-8} {\bibfield  {journal} {\bibinfo  {journal}
  {Journal of Low Temperature Physics}\ }\textbf {\bibinfo {volume} {209}},\
  \bibinfo {pages} {363--371} (\bibinfo {year} {2022})}\BibitemShut {NoStop}%
\bibitem [{\citenamefont {Dibert}\ \emph {et~al.}(2023)\citenamefont {Dibert},
  \citenamefont {Barry}, \citenamefont {Anderson}, \citenamefont {Benson},
  \citenamefont {Cecil}, \citenamefont {Chang}, \citenamefont {Fichman},
  \citenamefont {Karkare}, \citenamefont {Li}, \citenamefont {Natoli},
  \citenamefont {Pan}, \citenamefont {Rouble}, \citenamefont {Shirokoff},\ and\
  \citenamefont {Young}}]{dibert2023}%
  \BibitemOpen
  \bibfield  {author} {\bibinfo {author} {\bibfnamefont {K.~R.}\ \bibnamefont
  {Dibert}}, \bibinfo {author} {\bibfnamefont {P.~S.}\ \bibnamefont {Barry}},
  \bibinfo {author} {\bibfnamefont {A.~J.}\ \bibnamefont {Anderson}}, \bibinfo
  {author} {\bibfnamefont {B.~A.}\ \bibnamefont {Benson}}, \bibinfo {author}
  {\bibfnamefont {T.}~\bibnamefont {Cecil}}, \bibinfo {author} {\bibfnamefont
  {C.~L.}\ \bibnamefont {Chang}}, \bibinfo {author} {\bibfnamefont {K.~N.}\
  \bibnamefont {Fichman}}, \bibinfo {author} {\bibfnamefont {K.}~\bibnamefont
  {Karkare}}, \bibinfo {author} {\bibfnamefont {J.}~\bibnamefont {Li}},
  \bibinfo {author} {\bibfnamefont {T.}~\bibnamefont {Natoli}}, \bibinfo
  {author} {\bibfnamefont {Z.}~\bibnamefont {Pan}}, \bibinfo {author}
  {\bibfnamefont {M.}~\bibnamefont {Rouble}}, \bibinfo {author} {\bibfnamefont
  {E.}~\bibnamefont {Shirokoff}}, \ and\ \bibinfo {author} {\bibfnamefont
  {M.}~\bibnamefont {Young}},\ }\bibfield  {title} {\enquote {\bibinfo {title}
  {{Characterization of MKIDs for CMB Observation at 220 GHz With the South
  Pole Telescope}},}\ }\href {\doibase 10.1109/TASC.2023.3250386} {\bibfield
  {journal} {\bibinfo  {journal} {IEEE Transactions on Applied
  Superconductivity}\ }\textbf {\bibinfo {volume} {33}},\ \bibinfo {pages}
  {1--5} (\bibinfo {year} {2023})}\BibitemShut {NoStop}%
\bibitem [{\citenamefont {Barry}\ \emph {et~al.}(2022)\citenamefont {Barry},
  \citenamefont {Anderson}, \citenamefont {Benson}, \citenamefont {Carlstrom},
  \citenamefont {Cecil}, \citenamefont {Chang}, \citenamefont {Dobbs},
  \citenamefont {Hollister}, \citenamefont {Karkare}, \citenamefont {Keating},
  \citenamefont {Marrone}, \citenamefont {McMahon}, \citenamefont {Montgomery},
  \citenamefont {Pan}, \citenamefont {Robson}, \citenamefont {Rouble},
  \citenamefont {Shirokoff},\ and\ \citenamefont {Smecher}}]{barry2022}%
  \BibitemOpen
  \bibfield  {author} {\bibinfo {author} {\bibfnamefont {P.~S.}\ \bibnamefont
  {Barry}}, \bibinfo {author} {\bibfnamefont {A.}~\bibnamefont {Anderson}},
  \bibinfo {author} {\bibfnamefont {B.}~\bibnamefont {Benson}}, \bibinfo
  {author} {\bibfnamefont {J.~E.}\ \bibnamefont {Carlstrom}}, \bibinfo {author}
  {\bibfnamefont {T.}~\bibnamefont {Cecil}}, \bibinfo {author} {\bibfnamefont
  {C.}~\bibnamefont {Chang}}, \bibinfo {author} {\bibfnamefont
  {M.}~\bibnamefont {Dobbs}}, \bibinfo {author} {\bibfnamefont
  {M.}~\bibnamefont {Hollister}}, \bibinfo {author} {\bibfnamefont {K.~S.}\
  \bibnamefont {Karkare}}, \bibinfo {author} {\bibfnamefont {G.~K.}\
  \bibnamefont {Keating}}, \bibinfo {author} {\bibfnamefont {D.}~\bibnamefont
  {Marrone}}, \bibinfo {author} {\bibfnamefont {J.}~\bibnamefont {McMahon}},
  \bibinfo {author} {\bibfnamefont {J.}~\bibnamefont {Montgomery}}, \bibinfo
  {author} {\bibfnamefont {Z.}~\bibnamefont {Pan}}, \bibinfo {author}
  {\bibfnamefont {G.}~\bibnamefont {Robson}}, \bibinfo {author} {\bibfnamefont
  {M.}~\bibnamefont {Rouble}}, \bibinfo {author} {\bibfnamefont
  {E.}~\bibnamefont {Shirokoff}}, \ and\ \bibinfo {author} {\bibfnamefont
  {G.}~\bibnamefont {Smecher}},\ }\bibfield  {title} {\enquote {\bibinfo
  {title} {{Design of the SPT-SLIM Focal Plane: A Spectroscopic Imaging Array
  for the South Pole Telescope}},}\ }\href {\doibase
  10.1007/s10909-022-02843-4} {\bibfield  {journal} {\bibinfo  {journal}
  {Journal of Low Temperature Physics}\ }\textbf {\bibinfo {volume} {209}},\
  \bibinfo {pages} {879–888} (\bibinfo {year} {2022})}\BibitemShut {NoStop}%
\bibitem [{\citenamefont {Anderson}\ \emph {et~al.}(2022)\citenamefont
  {Anderson}, \citenamefont {Barry}, \citenamefont {Bender}, \citenamefont
  {Benson}, \citenamefont {Bleem}, \citenamefont {Carlstrom}, \citenamefont
  {Cecil}, \citenamefont {Chang}, \citenamefont {Crawford}, \citenamefont
  {Dibert}, \citenamefont {Dobbs}, \citenamefont {Fichman}, \citenamefont
  {Halverson}, \citenamefont {Holzapfel}, \citenamefont {Hryciuk},
  \citenamefont {Karkare}, \citenamefont {Li}, \citenamefont {Lisovenko},
  \citenamefont {Marrone}, \citenamefont {McMahon}, \citenamefont {Montgomery},
  \citenamefont {Natoli}, \citenamefont {Pan}, \citenamefont {Raghunathan},
  \citenamefont {Reichardt}, \citenamefont {Rouble}, \citenamefont {Shirokoff},
  \citenamefont {Smecher}, \citenamefont {Stark}, \citenamefont {Vieira},\ and\
  \citenamefont {Young}}]{anderson2022}%
  \BibitemOpen
  \bibfield  {author} {\bibinfo {author} {\bibfnamefont {A.~J.}\ \bibnamefont
  {Anderson}}, \bibinfo {author} {\bibfnamefont {P.}~\bibnamefont {Barry}},
  \bibinfo {author} {\bibfnamefont {A.~N.}\ \bibnamefont {Bender}}, \bibinfo
  {author} {\bibfnamefont {B.~A.}\ \bibnamefont {Benson}}, \bibinfo {author}
  {\bibfnamefont {L.~E.}\ \bibnamefont {Bleem}}, \bibinfo {author}
  {\bibfnamefont {J.~E.}\ \bibnamefont {Carlstrom}}, \bibinfo {author}
  {\bibfnamefont {T.~W.}\ \bibnamefont {Cecil}}, \bibinfo {author}
  {\bibfnamefont {C.~L.}\ \bibnamefont {Chang}}, \bibinfo {author}
  {\bibfnamefont {T.~M.}\ \bibnamefont {Crawford}}, \bibinfo {author}
  {\bibfnamefont {K.~R.}\ \bibnamefont {Dibert}}, \bibinfo {author}
  {\bibfnamefont {M.~A.}\ \bibnamefont {Dobbs}}, \bibinfo {author}
  {\bibfnamefont {K.}~\bibnamefont {Fichman}}, \bibinfo {author} {\bibfnamefont
  {N.~W.}\ \bibnamefont {Halverson}}, \bibinfo {author} {\bibfnamefont {W.~L.}\
  \bibnamefont {Holzapfel}}, \bibinfo {author} {\bibfnamefont {A.}~\bibnamefont
  {Hryciuk}}, \bibinfo {author} {\bibfnamefont {K.~S.}\ \bibnamefont
  {Karkare}}, \bibinfo {author} {\bibfnamefont {J.}~\bibnamefont {Li}},
  \bibinfo {author} {\bibfnamefont {M.}~\bibnamefont {Lisovenko}}, \bibinfo
  {author} {\bibfnamefont {D.}~\bibnamefont {Marrone}}, \bibinfo {author}
  {\bibfnamefont {J.}~\bibnamefont {McMahon}}, \bibinfo {author} {\bibfnamefont
  {J.}~\bibnamefont {Montgomery}}, \bibinfo {author} {\bibfnamefont
  {T.}~\bibnamefont {Natoli}}, \bibinfo {author} {\bibfnamefont
  {Z.}~\bibnamefont {Pan}}, \bibinfo {author} {\bibfnamefont {S.}~\bibnamefont
  {Raghunathan}}, \bibinfo {author} {\bibfnamefont {C.~L.}\ \bibnamefont
  {Reichardt}}, \bibinfo {author} {\bibfnamefont {M.}~\bibnamefont {Rouble}},
  \bibinfo {author} {\bibfnamefont {E.}~\bibnamefont {Shirokoff}}, \bibinfo
  {author} {\bibfnamefont {G.}~\bibnamefont {Smecher}}, \bibinfo {author}
  {\bibfnamefont {A.~A.}\ \bibnamefont {Stark}}, \bibinfo {author}
  {\bibfnamefont {J.~D.}\ \bibnamefont {Vieira}}, \ and\ \bibinfo {author}
  {\bibfnamefont {M.~R.}\ \bibnamefont {Young}},\ }\bibfield  {title} {\enquote
  {\bibinfo {title} {{SPT-3G+: mapping the high-frequency cosmic microwave
  background using kinetic inductance detectors}},}\ }in\ \href {\doibase
  10.1117/12.2629755} {\emph {\bibinfo {booktitle} {Millimeter, Submillimeter,
  and Far-Infrared Detectors and Instrumentation for Astronomy XI}}},\ Vol.\
  \bibinfo {volume} {12190},\ \bibinfo {editor} {edited by\ \bibinfo {editor}
  {\bibfnamefont {J.}~\bibnamefont {Zmuidzinas}}\ and\ \bibinfo {editor}
  {\bibfnamefont {J.-R.}\ \bibnamefont {Gao}}},\ \bibinfo {organization}
  {International Society for Optics and Photonics}\ (\bibinfo  {publisher}
  {SPIE},\ \bibinfo {year} {2022})\ p.\ \bibinfo {pages} {1219003}\BibitemShut
  {NoStop}%
\bibitem [{\citenamefont {Bandura}\ \emph {et~al.}(2016)\citenamefont
  {Bandura}, \citenamefont {Bender}, \citenamefont {Cliche}, \citenamefont
  {Haan}, \citenamefont {Dobbs}, \citenamefont {Gilbert}, \citenamefont
  {Griffin}, \citenamefont {Hsyu}, \citenamefont {Ittah}, \citenamefont {Mena},
  \citenamefont {Montgomery}, \citenamefont {Pinsonneault-Marotte},
  \citenamefont {Siegel}, \citenamefont {Smecher}, \citenamefont {Tang},
  \citenamefont {Vanderlinde},\ and\ \citenamefont {Whitehorn}}]{bandura2016}%
  \BibitemOpen
  \bibfield  {author} {\bibinfo {author} {\bibfnamefont {K.}~\bibnamefont
  {Bandura}}, \bibinfo {author} {\bibfnamefont {A.}~\bibnamefont {Bender}},
  \bibinfo {author} {\bibfnamefont {J.-F.}\ \bibnamefont {Cliche}}, \bibinfo
  {author} {\bibfnamefont {T.}~\bibnamefont {Haan}}, \bibinfo {author}
  {\bibfnamefont {M.}~\bibnamefont {Dobbs}}, \bibinfo {author} {\bibfnamefont
  {A.}~\bibnamefont {Gilbert}}, \bibinfo {author} {\bibfnamefont
  {S.}~\bibnamefont {Griffin}}, \bibinfo {author} {\bibfnamefont
  {G.}~\bibnamefont {Hsyu}}, \bibinfo {author} {\bibfnamefont {D.}~\bibnamefont
  {Ittah}}, \bibinfo {author} {\bibfnamefont {J.}~\bibnamefont {Mena}},
  \bibinfo {author} {\bibfnamefont {J.}~\bibnamefont {Montgomery}}, \bibinfo
  {author} {\bibfnamefont {T.}~\bibnamefont {Pinsonneault-Marotte}}, \bibinfo
  {author} {\bibfnamefont {S.}~\bibnamefont {Siegel}}, \bibinfo {author}
  {\bibfnamefont {G.}~\bibnamefont {Smecher}}, \bibinfo {author} {\bibfnamefont
  {Q.}~\bibnamefont {Tang}}, \bibinfo {author} {\bibfnamefont {K.}~\bibnamefont
  {Vanderlinde}}, \ and\ \bibinfo {author} {\bibfnamefont {N.}~\bibnamefont
  {Whitehorn}},\ }\bibfield  {title} {\enquote {\bibinfo {title} {{ICE}: A
  scalable, low-cost {FPGA}-based telescope signal processing and networking
  system},}\ }\href {\doibase 10.1142/S2251171716410051} {\bibfield  {journal}
  {\bibinfo  {journal} {Journal of Astronomical Instrumentation}\ }\textbf
  {\bibinfo {volume} {05}} (\bibinfo {year} {2016}),\
  10.1142/S2251171716410051}\BibitemShut {NoStop}%
\bibitem [{\citenamefont {Mani}()}]{cryoelec}%
  \BibitemOpen
  \bibfield  {author} {\bibinfo {author} {\bibfnamefont {H.}~\bibnamefont
  {Mani}},\ }\href@noop {} {\enquote {\bibinfo {title} {{CryoElec Low Noise
  Amplifier}},}\ }\bibinfo {type} {{CryoElec}}\BibitemShut {NoStop}%
\bibitem [{\citenamefont {Goldie}\ and\ \citenamefont
  {Withington}(2012)}]{goldie2013}%
  \BibitemOpen
  \bibfield  {author} {\bibinfo {author} {\bibfnamefont {D.~J.}\ \bibnamefont
  {Goldie}}\ and\ \bibinfo {author} {\bibfnamefont {S.}~\bibnamefont
  {Withington}},\ }\bibfield  {title} {\enquote {\bibinfo {title}
  {Non-equilibrium superconductivity in quantum-sensing superconducting
  resonators},}\ }\href {\doibase 10.1088/0953-2048/26/1/015004} {\bibfield
  {journal} {\bibinfo  {journal} {Superconductor Science and Technology}\
  }\textbf {\bibinfo {volume} {26}},\ \bibinfo {pages} {015004} (\bibinfo
  {year} {2012})}\BibitemShut {NoStop}%
\bibitem [{\citenamefont {McCarrick}\ \emph {et~al.}(2014)\citenamefont
  {McCarrick}, \citenamefont {Flanigan}, \citenamefont {Jones}, \citenamefont
  {Johnson}, \citenamefont {Ade}, \citenamefont {Araujo}, \citenamefont
  {Bradford}, \citenamefont {Cantor}, \citenamefont {Che}, \citenamefont {Day},
  \citenamefont {Doyle}, \citenamefont {Leduc}, \citenamefont {Limon},
  \citenamefont {Luu}, \citenamefont {Mauskopf}, \citenamefont {Miller},
  \citenamefont {Mroczkowski}, \citenamefont {Tucker},\ and\ \citenamefont
  {Zmuidzinas}}]{McCarrick2014}%
  \BibitemOpen
  \bibfield  {author} {\bibinfo {author} {\bibfnamefont {H.}~\bibnamefont
  {McCarrick}}, \bibinfo {author} {\bibfnamefont {D.}~\bibnamefont {Flanigan}},
  \bibinfo {author} {\bibfnamefont {G.}~\bibnamefont {Jones}}, \bibinfo
  {author} {\bibfnamefont {B.~R.}\ \bibnamefont {Johnson}}, \bibinfo {author}
  {\bibfnamefont {P.}~\bibnamefont {Ade}}, \bibinfo {author} {\bibfnamefont
  {D.}~\bibnamefont {Araujo}}, \bibinfo {author} {\bibfnamefont
  {K.}~\bibnamefont {Bradford}}, \bibinfo {author} {\bibfnamefont
  {R.}~\bibnamefont {Cantor}}, \bibinfo {author} {\bibfnamefont
  {G.}~\bibnamefont {Che}}, \bibinfo {author} {\bibfnamefont {P.}~\bibnamefont
  {Day}}, \bibinfo {author} {\bibfnamefont {S.}~\bibnamefont {Doyle}}, \bibinfo
  {author} {\bibfnamefont {H.}~\bibnamefont {Leduc}}, \bibinfo {author}
  {\bibfnamefont {M.}~\bibnamefont {Limon}}, \bibinfo {author} {\bibfnamefont
  {V.}~\bibnamefont {Luu}}, \bibinfo {author} {\bibfnamefont {P.}~\bibnamefont
  {Mauskopf}}, \bibinfo {author} {\bibfnamefont {A.}~\bibnamefont {Miller}},
  \bibinfo {author} {\bibfnamefont {T.}~\bibnamefont {Mroczkowski}}, \bibinfo
  {author} {\bibfnamefont {C.}~\bibnamefont {Tucker}}, \ and\ \bibinfo {author}
  {\bibfnamefont {J.}~\bibnamefont {Zmuidzinas}},\ }\bibfield  {title}
  {\enquote {\bibinfo {title} {Horn‐coupled, commercially‐fabricated
  aluminum lumped‐element kinetic inductance detectors for millimeter
  wavelengths},}\ }\href {\doibase 10.1063/1.4903855} {\bibfield  {journal}
  {\bibinfo  {journal} {Review of Scientific Instruments}\ }\textbf {\bibinfo
  {volume} {85}},\ \bibinfo {pages} {123117} (\bibinfo {year}
  {2014})}\BibitemShut {NoStop}%
\end{thebibliography}%

\appendix

\section{Resonator \& circuit model parameters}\label{app:model_params}

\begin{table*}[htbp]
  \centering
  \caption{Resonator model parameters}
  \label{tab:model_params}
  \begin{ruledtabular}
  \begin{tabular}{@{} 
      l   
      l   
      c   
      l   
      p{6cm} 
    @{}}
    Symbol & Description                & Value               & Units        & Notes \\
    \midrule
    \multicolumn{4}{@{}l}{\bfseries Model parameters} \\ 
         	& Inductor material & Aluminum &  &  \\
    $l$    & Inductor length                      & $8.33\times10^{-3}$   & m      &       \\
    $w$    & Inductor width                       & $2.00\times10^{-6}$   & m       &      \\
    $t$    & Inductor thickness                   & $3.00\times10^{-8}$   & m        &     \\
    $P_{opt}$ 
           & Optical load          & $2.50\times10^{-15}$  & W  & used to compute quasiparticle density    \\
               $T$    & Operating temperature       & $0.12$                & K & used to compute quasiparticle density \\
    $I_*$  & Nonlinear scaling current            & $0.94\times10^{-3}$   & A             \\

\multirow[c]{2}{*}{$\Delta_0$}  & \multirow[c]{2}{*}{Gap energy (zero-temperature)}  & \multirow[c]{2}{*}{$1.76 k_B T_c$ }  & \multirow[c]{2}{*}{J} & for $\Delta(T)$ we use the numerical approximation from Ref. \onlinecite{gao2008} \\
    	$\sigma_N$  & Normal-state conductivity  & $1.25\times10^{7}$  & $\mathrm{(\Omega m)^{-1}}$ & as measured in Ref. \onlinecite{McCarrick2014} for a 20 nm Al film  \\
    $T_c$       & Critical temperature             & $1.20$                  & K         & \\
    $N_0$       & Single-spin density of states     & $2.76\times10^{29}$     & $\mathrm{\mu m^{-3}\,J^{-1}}$  & \\

    \multicolumn{4}{@{}l}{\bfseries Circuit components} \\ 
    \multirow[c]{2}{*}{$R$}    & \multirow[c]{2}{*}{Resonator real impedance }                 & \multirow[c]{2}{*}{$7.2407\times10^{-4}$} & \multirow[c]{2}{*}{$\Omega$}  & from inductor geometry and complex conductivity \\
    \multirow[c]{2}{*}{$L_k$ } & \multirow[c]{2}{*}{Zero-current kinetic inductance }         & \multirow[c]{2}{*}{ $1.2797\times10^{-8}$} & \multirow[c]{2}{*}{ H}   & from inductor geometry and complex conductivity  \\
    $L_g$  & Geometric inductance        & $2.9860\times10^{-8}$ & H   &  $\alpha_k = 0.3$        \\
    $C$    & Resonator capacitance       & $3.70\times10^{-13}$  & F    &        \\
    $C_c$  & Coupling capacitance        & $9.10\times10^{-15}$  & F    &        \\
    $Z_{LNA}$ & LNA input impedance & 50 & $\Omega$ & \\
    $R_1$, $R_2$, $R_3$ & resistances in last-stage attenuator & 61.1. 247.5, 61.1 & $\Omega$ & a 20 dB attenuator

  \end{tabular}
  \end{ruledtabular}
\end{table*}

\end{document}